\documentclass[12pt,english,american]{article}
\usepackage[T1]{fontenc}
\usepackage[latin9]{inputenc}
\usepackage{geometry}
\usepackage{color}
\usepackage{babel}
\usepackage{amsthm}
\usepackage{amstext}
\usepackage{amssymb}
\usepackage{enumitem}
\usepackage{amsmath}
\usepackage{simplewick}
\usepackage[unicode=true,pdfusetitle,
 bookmarks=true,bookmarksnumbered=false,bookmarksopen=false,
 breaklinks=true,pdfborder={0 0 0},backref=false,colorlinks=true]
 {hyperref}
\hypersetup{
 linkcolor=blue,citecolor=blue, urlcolor=blue}

\makeatletter

\usepackage{enumitem}		
\theoremstyle{plain}

\theoremstyle{definition}

\theoremstyle{plain}

\theoremstyle{remark}

\theoremstyle{plain}

\theoremstyle{plain}

\usepackage{txfonts,refstyle,xcolor}

\newref{con}{name = Conjecture\ }
\newref{prop}{name = Proposition\ }
\newref{def}{name = Definition\ }
\newref{sec}{name = Section\ }
\newref{sub}{name = Section\ }
\newref{thm}{name = Theorem\ }
\newref{lem}{name = Lemma\ }
\newref{cor}{name = Corollary\ }
\newref{fig}{name = Figure\ }

\usepackage{bbm}

\usepackage[all]{xy}

\newcommand{\xyR}[1]{%
     \makeatletter
     \xydef@\xymatrixrowsep@{#1}
     \makeatother
}

\newcommand{\xyC}[1]{%
     \makeatletter
     \xydef@\xymatrixcolsep@{#1}
     \makeatother
}

\newcommand{\ncol}[1]{\color{normalcolor}}

\makeatother

\addto\captionsamerican{\renewcommand{\corollaryname}{Corollary}}
\addto\captionsamerican{\renewcommand{\definitionname}{Definition}}
\addto\captionsamerican{\renewcommand{\lemmaname}{Lemma}}
\addto\captionsamerican{\renewcommand{\propositionname}{Proposition}}
\addto\captionsamerican{\renewcommand{\remarkname}{Remark}}
\addto\captionsamerican{\renewcommand{\theoremname}{Theorem}}
\addto\captionsenglish{\renewcommand{\corollaryname}{Corollary}}
\addto\captionsenglish{\renewcommand{\definitionname}{Definition}}
\addto\captionsenglish{\renewcommand{\lemmaname}{Lemma}}
\addto\captionsenglish{\renewcommand{\propositionname}{Proposition}}
\addto\captionsenglish{\renewcommand{\remarkname}{Remark}}
\addto\captionsenglish{\renewcommand{\theoremname}{Theorem}}
\providecommand{\corollaryname}{Corollary}
\providecommand{\definitionname}{Definition}
\providecommand{\lemmaname}{Lemma}
\providecommand{\propositionname}{Proposition}
\providecommand{\remarkname}{Remark}
\providecommand{\theoremname}{Theorem}

\newcommand{\tiga}{\ga_0}
\newcommand{\siz}{\ka}
\newcommand{\ba}{\,|\,}
\newcommand{\nF}{V}

\newcommand{\s}{\mathrm{s}}
\newcommand{\Gn}{G_{1,n}}
\newcommand{\Hm}{G_{2,m}}
\newcommand{\tin}{\ti n}
\newcommand{\tim}{\ti m}
\newcommand{\ovGtin}{\ov G_{2,\ti n} }
\newcommand{\ovHtim}{\ov G_{1,\ti m} }
\newcommand{\Gtin}{G_{2,\ti n}}
\newcommand{\Htim}{G_{1,\ti m}}
\newcommand{\mcD}{\mathcal D}
\newcommand{\B}{C_0^{2}}
\newcommand{\tic}{c}
\newcommand{\sym}{\mathrm{sym} }
\newcommand{\PS}{S}
\newcommand{\maxi}{\mathrm{max}}
\newcommand{\fib}{\mathrm{fi}}
\newcommand{\free}{\mathrm{fr}}
\newcommand{\I}{\mathrm I}
\newcommand{\ain}{\mathrm a}
\newcommand{\cin}{\mathrm c}
\newcommand{\hc}{\mathrm{h.c.}}
\newcommand{\mcA}{\mathcal A}
\newcommand{\mcB}{\mathcal B}
\newcommand{\mcC}{\mathcal C}
\newcommand{\el}{\mathrm e}

\newcommand{\kas}{\kappa_{*}}

\newcommand{\sig}{\rho}
\newcommand{\ci}{i_*}

\newcommand{\pha}{\phantom}
\newcommand{\cvv}{\check{v}_{\alf} }
\newcommand{\cH}{\check H}
\newcommand{\unk}{\un k}
\newcommand{\unr}{\un r}
\newcommand{\hk}{\hat k }    
\newcommand{\hr}{\hat r } 
\newcommand{\chk}{\check k } 
\newcommand{\chr}{\check r } 
\newcommand{\uhk}{\un{\hat k}  } 
\newcommand{\uhr}{\un{\hat r}  }  
\newcommand{\uchk}{\un{\check k}  }  
\newcommand{\uchr}{\un{\check r}  }  
\newcommand{\sip}{\si'}
\newcommand{\pho}{\mathrm{f}}
\newcommand{\alf}{\overline{\al}}
\newcommand{\nr}{\reta}
\newcommand{\nn}{\eta}
\newcommand{\neta}{\eta}
\newcommand{\reta}{\hat\eta}

\newcommand{\bb}{a}

\newcommand{\vv}{v_{\alf}}

\newcommand{\slim}{\lim}    

\newcommand{\g}{\la }

\newcommand{\ti}{\tilde}

\newcommand{\un}{\underline}

\newcommand{\vac}{\Omega}
\newcommand{\Om}{\Omega}
\newcommand{\ga}{\gamma}

\newcommand{\ka}{\kappa}

\newcommand{\be}{\beta}

\newcommand{\pa}{\partial}

\newcommand{\ov}{\overline}
\newcommand{\vp}{\varphi}
\newcommand{\mfh}{\mathfrak{h}}

\newcommand{\eps}{\varepsilon}
\newcommand{\de}{\delta}

\newcommand{\De}{\Delta}

\newcommand{\si}{\sigma}
\newcommand{\ph}{\phantom}
\newcommand{\h}{\fr{1}{2}}
\newcommand{\nat}{\mathbb{N}}
\newcommand{\hil}{\mathcal{H}}
\newcommand{\om}{\omega}

\newcommand{\supp}{\mathrm{supp}}
\newcommand{\fr}[2]{\frac{#1}{#2}}
\newcommand{\al}{\alpha}
\newcommand{\real}{\mathbb{R}}

\newcommand{\la}{\lambda}
\newcommand{\non}{\nonumber}
\newcommand{\Ga}{\Gamma}
\newcommand{\half}{\fr{1}{2}}
\newcommand{\lan}{\langle}
\newcommand{\ran}{\rangle}

\def\proof{\noindent{\bf Proof. }}
\def\qed{$\Box$\medskip}

\newtheorem{theoreme}{Theorem } [section]
\newtheorem{proposition}[theoreme]{Proposition}
\newtheorem{lemma}[theoreme]{Lemma}
\newtheorem{definition}[theoreme]{Definition}
\newtheorem{corollary}[theoreme]{Corollary}
\newtheorem{remark}[theoreme]{Remark}
\newtheorem{example}[theoreme]{Example}
\newtheorem{criterion}[theoreme]{Criterion}

\newcommand{\beq}{\begin{equation}}
\newcommand{\eeq}{\end{equation}}
\newcommand{\beqa}{\begin{eqnarray}}
\newcommand{\eeqa}{\end{eqnarray}}
\newcommand{\ben}{\begin{arabicenumerate}}
\newcommand{\een}{\end{arabicenumerate}}
\newcommand{\bex}{\begin{example}}
\newcommand{\eex}{\end{example}}
\newcommand{\ber}{\begin{remark}}
\newcommand{\eer}{\end{remark}}
\newcommand{\bec}{\begin{corollary}}
\newcommand{\eec}{\end{corollary}}
\newcommand{\bep}{\begin{proposition}}
\newcommand{\eep}{\end{proposition}}
\newcommand{\becr}{\begin{criterion}}
\newcommand{\eecr}{\end{criterion}}

\def\bel{\begin{lemma}}
\def\eel{\end{lemma}}
\def\bet{\begin{theoreme}}
\def\eet{\end{theoreme}}
\def\bed{\begin{definition}}
\def\eed{\end{definition}}

\begin{document}
\title{Coulomb scattering in the massless Nelson model I. Foundations of two-electron scattering} 
 \author{W. Dybalski and A. Pizzo}


\maketitle

\abstract 
We construct two-electron scattering states and verify their tensor product structure in the infrared-regular massless Nelson model. 
The proof follows the lines of Haag-Ruelle scattering theory:  Scattering state approximants are defined with the help of  two 
time-dependent renormalized creation operators of the electrons acting on the vacuum.  They depend on ground state wave functions of the 
(single-electron) fiber Hamiltonians with infrared cut-off. Convergence of these approximants as $t\to \infty$ is shown  with the help of 
Cook's method combined with a non-stationary phase argument. Removal of the infrared cut-off in the limit $t\to \infty$  requires  sharp estimates
on the derivatives of these ground state wave functions w.r.t. electron and photon momenta, with mild dependence on the infrared cut-off.
These key estimates, which carry information about the localization of electrons in space, are obtained in a companion paper with the help of iterative analytic perturbation theory. Our results hold in the weak coupling
regime.

\section{Introduction and results} \label{intro-and-results}
\setcounter{equation}{0}

The last two decades have witnessed substantial progress in the mathematical understanding of scattering 
of light and matter  in the framework of non-relativistic QED. These advances provided
a rigorous foundation for  the  physical description of the Compton \cite{CFP07,CFP09,Pi03,Pi05,FGS04} and 
Rayleigh scattering \cite{DG99,DG04,FGS02},
involving one massive particle (the `electron') and many massless excitations (`photons'). 
However, the case of Coulomb scattering, i.e., collisions of two electrons in the presence of photons, 
has remained outside of the scope of these investigations. This is a serious gap in our understanding,
given the tremendous importance of this scattering process, ranging from Rutherford's discovery of the structure
of atoms, to modern high-energy physics experiments.  This paper is a first
part of a larger investigation, whose goal is to put this important process on rigorous grounds.

A general framework  of scattering  theory  for several electrons
in models of non-relativistic QED, which has its roots in  Haag-Ruelle scattering theory \cite{Ha58,Ru62},  
was known to experts already more than three decades ago \cite{Fr73.1,Al73}. However,  a 
complete  construction of  scattering states was only possible in the  presence
of a fixed infrared cut-off (or non-zero photon mass), due to  severe infrared- and infraparticle problems.  
These difficulties were overcome at the single-electron level with the help of a 
novel multiscale technique \cite{Pi03,Pi05}, but the case of several electrons remained open to date. 
In this paper we provide a construction of two-electron scattering states in the
massless, infrared-regular Nelson model. Although the infraparticle problem does 
not arise in this situation, the infrared structure of the model is non-trivial and
requires major refinements of the multiscale technique. To clarify the origin of these
new  difficulties  and describe our  methods to tackle them, let us now  explain in non-technical 
terms the main steps of our analysis.        

Let $H$ be the Hamiltonian of the translationally invariant, infrared-regular, massless Nelson model stated in (\ref{full-Nelson-model}) below. 
It describes second-quantized  non-relativistic particles, which we  call electrons though they  obey the Bose statistics,
interacting with massless scalar bosons, which we call photons.
This Hamiltonian is a self-adjoint operator on the physical Hilbert space $\hil$ which is the tensor product of the electron
Fock space $\Ga(\mfh_{\el})$ and the photon Fock space $\Ga(\mfh_{\pho})$. 
As $H$ preserves the number of electrons, we can restrict it to the one-electron subspace obtaining  the Hamiltonian 
$H^{(1)}$. This Hamiltonian  has
the standard decomposition into the fiber Hamiltonians at fixed total momentum $H^{(1)}_P$, which are operators on the Fock space. 
In the infrared-regular case ($1/2\geq \alf>0$ in (\ref{form-factor-formula}) below), the fiber Hamiltonains have (normalized) ground states 
$\psi_{P}$, (for $P$ in some ball $\PS$ centered at zero),  corresponding to eigenvalues $E_P$.  Superpositions of such ground states of the form
\beqa
\psi_h:=\Pi^*\int^{\oplus}d^3P\, h(P)\psi_P, \quad h\in C_0^{2}(\PS), \label{intro-single-electron}
\eeqa
give physical single-electron states in $\hil$. (Here $\Pi^*$ is the standard identification between the fiber picture and the physical picture). We note that the time evolution of $\psi_h$ is given by
\beqa
e^{-iHt}\psi_h=\psi_{h_t}, \label{single-electron-evolution}
\eeqa
where $h_t(P):=e^{-iE_Pt}h(P)$. In heuristic terms, 
to construct a scattering state of two electrons one has to `multiply' two single-electron states in such a way that the result 
is a vector in $\hil$.  Haag-Ruelle scattering theory is, in essence, a prescription to perform such a multiplication. 
In the context of the Nelson model  one could attempt to implement this construction in the following way:
Let $\{f^m_{P}\}_{m\in \nat_0}$ be the $m$-photon components of $\psi_{P}$. Then (\ref{intro-single-electron}) can be rewritten
as follows
\beqa
\psi_h=\sum_{m=0}^{\infty}\fr{1}{\sqrt{m!}}\int d^3p\,d^{3m}k\,h(p) f^{m}_{p}(k_1,\ldots, k_m)\bb^*(k_1)\ldots \bb^*(k_m) \, 
\nn^*(p- \unk)\vac, 
\eeqa
where $\vac\in \hil$ is the vacuum vector, $\nn^*(p)$, $\bb^*(k)$ are the electron and photon creation operators
and $\unk:=k_1+\cdots+k_m$. This suggests the following definition of the \emph{renormalized creation operator}
\beqa
\nr^*(h):=\sum_{m=0}^{\infty}\fr{1}{\sqrt{m!}}\int d^3p\,d^{3m}k\,h(p) f^{m}_{p}(k_1,\ldots,k_m)\bb^*(k_1)\ldots \bb^*(k_m)\nn^*(p- \unk),
\eeqa  
which creates the physical single-electron state $\psi_h$ from the vacuum\footnote{We found this definiton of the renormalized
creation operator  and a construction of scattering states of several electrons in the Nelson model with a fixed infrared cut-off in unpublished 
notes of J. Fr\"ohlich \cite{Fr73.1}. Such renormalized creation operators  appear also in a  work of S.~Albeverio \cite{Al73}.  }. 
Now, given $h_1,h_2\in C_0^{2}(\PS)$ with disjoint (velocity) supports,  the scattering state $\ti\Psi^+_{h_1,h_2}$   
describing an asymptotic configuration of two independent electrons  $\psi_{h_1}$, $\psi_{h_2}$  is given by
\beqa
\ti\Psi^+_{h_1,h_2} =\slim_{t\to\infty}e^{iHt}\nr^*(h_{1,t})\nr^*(h_{2,t})\vac,  
\label{idealized-scattering-states}
\eeqa
if the approximating vectors on the r.h.s. are well defined and the limit exists. 

Unfortunately,  regularity properties of the $m$-photon  components $f^{m}_{P}$ of $\psi_{P}$, which control
the rate of convergence in (\ref{idealized-scattering-states}), are difficult to obtain due to the fact that the eigenvalue 
$E_P$ is located at the bottom
of the  continuous spectrum of $H^{(1)}_P$.  Therefore, we will not construct scattering states with the help of  
formula~(\ref{idealized-scattering-states}), but instead we will use more tractable approximating sequences: Let $H^{(1)}_{P,\si}$
be the fiber Hamiltonians with the infrared cut-off $\si$, defined precisely in (\ref{infrared-cut-off-Hamiltonian}). 
Their ground states $\psi_{P,\si}$, corresponding to eigenvalues $E_{P,\si}$, satisfy
\beqa
\slim_{\si\to 0}\psi_{P,\si}=\psi_P.
\eeqa
Let $\{f^{m}_{P,\si}\}_{m\in\nat_0}$ be the  $m$-photon  components of $\psi_{P,\si}$.  The   renormalized
creation operators with the infrared cut-off $\si$ have the form
\beqa
\nr^*_{\si}(h):=\sum_{m=0}^{\infty}\fr{1}{\sqrt{m!}}\int d^3p\,d^{3m}k\,h(p) f^{m}_{p,\si}(k_1,\ldots,k_m)\bb^*(k_1)\ldots \bb^*(k_m)\nn^*(p- \unk).
\eeqa
We will use them to construct the scattering states as follows
\beqa
\Psi^+_{h_1,h_2} :=\slim_{t\to\infty} e^{iHt}\nr_{\si_t}^*(h_{1,t})\nr_{\si_t}^*(h_{2,t})\Om, \label{cut-off-scattering-states}
\eeqa
where    $\si_t=\siz/t^{\ga}$,  $\ga>4$ and $\ka>0$ is the ultraviolet cut-off. The existence of this limit and its properties, which allow to interpret it
as an asymptotic configuration of two electrons, constitute the main result of this paper (Theorem~\ref{main-result-scattering} below).
The main steps of the proof, which rely on Cook's method combined with non-stationary phase analysis, are given in Section~\ref{main-steps}.
We conjecture that the limit (\ref{idealized-scattering-states}) also exists and coincides with (\ref{cut-off-scattering-states}). Supporting
evidence for this conjecture comes from algebraic quantum field theory, where the existence of scattering states in the infrared-regular 
situation has been proven without introducing cut-offs \cite{Dy05}. We do not expect any problems with generalizing our result to
an arbitrary number of  electrons or with changing the statistics of  our `electrons'  from Bose to Fermi. However, our aim
here is not to cover the most general situation, but rather to prepare grounds for our future investigation concerning 
Coulomb scattering in the infrared-singular case. The additional complications coming from the presence of infrared photon clouds
should first be tackled in the case of two electrons.

Let us conclude this introductory discussion with some more technical remarks. We point out that
the approximating sequence (\ref{cut-off-scattering-states}) has important  advantages over (\ref{idealized-scattering-states}): 
Due to the fact that $E_{P,\si}$ is an isolated eigenvalue, the relevant regularity properties of the eigenvectors $\psi_{P,\si}$ and
their $m$-photon components $f^{m}_{P,\si}$ can be obtained using the analytic perturbation theory.  Their behaviour in the
limit $\si\to 0$ can be studied with the help of the iterative multiscale technique. As  this analysis is presented in a separate 
paper \cite{DP12}, it suffices to indicate here the new spectral difficulties encountered at the level of two-electron scattering.   

We recall that  already in  \cite{Fr73, Fr74.1} a formula
for $f^{m}_{P,\si}$ was derived, which in the case of $m=1$ has the form 
\beqa
 f^{1}_{P,\si}(k) = -\lan \Om,\bigg\{   \fr{1}{H^{(1)}_{P-k,\si}-E_{P,\si}+|k|}
\vv^{\si}(k)\bigg\} \psi_{P,\si}\ran \label{Froehlich-formula},
\eeqa
where the form-factor $\vv^{\si}$ is given by (\ref{IR-cut-off-propagator}). It is not difficult to obtain from this formula that
\beqa
|f^{1}_{P,\si}(k)|\leq \fr{c \vv^{\si}(k)} {|k|}, \label{photon-components-standard}
\eeqa
where the constant $c$ is independent of $\si$. However, to prove convergence of (\ref{cut-off-scattering-states}) via
the non-stationary phase method, we also need  bounds on derivatives $\pa_{P^i} f^{m}_{P,\si}$, $\pa_{P^i}\pa_{P^j} f^{m}_{P,\si}$
with mild dependence on $\si$. To indicate various difficulties that have to be tackled to obtain such bounds, let us differentiate
(\ref{Froehlich-formula}) w.r.t. $P$:
\beqa
\pa_{P^i} f^{1}_{P,\si}(k)&=&\lan \Om,\bigg\{   \fr{1}{H^{(1)}_{P-k,\si}-E_{P,\si}+|k|}\big( (P-k-P_{\pho})-\nabla E_{P,\si}  \big)\fr{1}{H^{(1)}_{P-k,\si}-E_{P,\si}+|k|}\vv^{\si}(k)\bigg\} \psi_{P,\si}\ran\non\\
& &-\lan \Om,\bigg\{   \fr{1}{H^{(1)}_{P-k,\si}-E_{P,\si}+|k|} \vv^{\si}(k)\bigg\}\pa_{P^i} \psi_{P,\si}\ran, \label{intro-derivative}
\eeqa
where $P_{\pho}$ is the photon momentum operator. As for the second term on the r.h.s. of (\ref{intro-derivative}) the main difficulty
is to find an estimate on  the vector $\pa_{P^i} \psi_{P,\si}$ with suitable dependence on $\si$. We recall that the
existing bounds from \cite{Pi03} give only H\"older continuity of $P\mapsto \psi_{P,\si}$, uniformly in $\si$. By a refinement of
the argument from \cite{Pi03},  we obtain in \cite{DP12} the following bound, stated precisely in Proposition~\ref{preliminaries-on-spectrum} below, 
\beqa
\|\pa_{P^i} \psi_{P,\si}\|\leq \fr{c}{\si^{\de_{\g_0}}},
\eeqa 
where $c$ is independent of $\si$, $\g_0>0$ is the maximal admissible value of the coupling constant and the function $\g_0\mapsto \de_{\g_0}$ satisfies $\lim_{\g_0\to 0}\de_{\g_0}=0$. Although the bound is not uniform in $\si$, such mild dependence on the infrared cut-off is sufficient for our purposes.
Denoting by $II$ the second term on the r.h.s. of (\ref{intro-derivative}), and proceeding as in the derivation of (\ref{photon-components-standard}),
we obtain 
\beqa
|II|\leq \fr{1}{\si^{\de_{\g_0}}}  \fr{c \vv^{\si}(k)} {|k|}.
\eeqa

As for the first term on the r.h.s. of (\ref{intro-derivative}), which we denote by $I$, a crude estimate gives
\beqa
|I|\leq \fr{c \vv^{\si}(k)} {|k|^2}\leq \fr{1}{\si}\fr{c \vv^{\si}(k)} {|k|},
\eeqa  
due to the presence of the additional resolvent and the support properties of $\vv^{\si}$. This bound does not suffice for the purpose of
constructing scattering states. With the help of the multiscale analysis we improve it to
\beqa
|I|\leq \fr{1}{\si^{\de_{\g_0}}}\fr{c \vv^{\si}(k)} {|k|},
\eeqa 
which is only slightly worse than (\ref{photon-components-standard}). Altogether we get
\beqa
|\pa_{P^i} f^{1}_{P,\si}(k)|\leq \fr{1}{\si^{\de_{\g_0}}}\fr{c \vv^{\si}(k)} {|k|}
\eeqa
and an analogous bound for the second derivative. These bounds, and their counterparts for  $f^{m}_{P,\si}$, $m\geq 1$,
constitute the main result of \cite{DP12}, stated also in Theorem~\ref{main-theorem-spectral} below. 

This paper is organized as follows: In Subsection~\ref{model-sub} we recall the definition of  the Nelson model with many electrons. 
In Subsection~\ref{spectral-sub} we state the relevant results concerning spectral theory, which are proven in a separate paper \cite{DP12}. 
In Subsection~\ref{scatter-sub} we state the main result of the present paper which concern scattering theory of two electrons in the 
infrared-regular massless Nelson model. Section~\ref{main-steps} presents the main steps of the proof and in the subsequent sections
we provide the necessary ingredients.   Section~\ref{vacuum-expectation-values} is devoted to vacuum expectation values of the 
renormalized creation operators. In Section~\ref{non-stationary-phase}
decay properties of these vacuum expectation values are derived with the help of the method of non-stationary phase. Important input
in this section are our spectral results proven in \cite{DP12} and summarized in Subsection~\ref{spectral-sub}. The more technical part
of our discussion is postponed to appendices.

\vspace{0.5cm}

\noindent{\bf Acknowledgment:}  The authors are  grateful to J. Fr\"ohlich for the unpublished notes that have
inspired this paper. We also thank the Hausdorff Research Institute for
Mathematics, Bonn, for hospitality at final stages of this work. W.D.  acknowledges hospitality of the University
of California Davis, where this collaboration started.  

A.P. is supported by the NSF grant \#DMS-0905988.
W.D. is supported by the German Research Foundation (DFG) within the
grant SP181/25--2 and stipend DY107/1--1. Moreover, W.D. would like to acknowledge the support of 
the Danish Council for Independent Research, grant no. 09-065927 "Mathematical Physics", and of the Lundbeck Foundation.

\subsection{The model}\label{model-sub}

We consider an interacting system of massive spinless bosons, which we will call `electrons',  and massless
spinless bosons, which we will call `photons'.   
Let $\mfh_{\el}:=L^2(\real^3,d^3p)$ and $\mfh_{\pho}:=L^2(\real^3,d^3k)$ be the single-electron and 
single-photon spaces, respectively, and let $\Ga(\mfh_{\el})$ and $\Ga(\mfh_{\pho})$ be the corresponding
symmetric Fock spaces. The (improper) creation and annihilation operators on $\Ga(\mfh_{\el})$ (resp. $\Ga(\mfh_{\pho})$)
will be denoted by $\nn^*(p),\nn(p)$ (resp. $\bb^*(k), \bb(k)$). They satisfy the canonical commutation relations:
\beqa
& &[\nn(p),\nn^*(p')]=\de(p-p'),\quad [\nn(p),\nn(p')]=[\nn^*(p),\nn^*(p')]=0, \\
& &[\bb(k),\bb^*(k')]=\de(k-k'),\quad \,\,\, [\bb(k),\bb(k')]=[\bb^*(k),\bb^*(k')]=0.
\eeqa
The free Hamiltonians of electrons and photons are given by
\beqa
H_{\el}:=\int d^3p\, \Om(p)\nn^*(p)\nn(p),\quad H_{\pho}:=\int d^3k\, \om(k) a^*(k)a(k),
\eeqa
where $\Om(p)=\fr{p^2}{2}$ and $\om(k)=|k|$. We recall that these operators 
are essentially self-adjoint on $\mcC_{\el}$, $\mcC_{\pho}$, respectively, where $\mcC_{\el/\pho}\subset \Ga(\mfh_{\el/\pho})$
are  dense subspaces consisting of finite linear combinations of symmetrized tensor products of elements of $C_0^{\infty}(\real^3)$.

The physical Hilbert space of our system is $\hil:= \Ga(\mfh_{\el})\otimes  \Ga(\mfh_{\pho})$
and we will follow the standard convention to denote operators of the form $A\otimes 1$ and $1\otimes B$ by $A$ and
$B$, respectively. The Hamiltonian describing the free evoluton of the composite system of electrons and photons is given by
\beqa
H_{\free}:=H_{\el}+H_{\pho}
\eeqa
and it is essentially self-adjoint on $\mcC:=\mcC_{\el}\otimes \mcC_{\pho}$. Now let us introduce the interaction  between 
electrons and photons. Let $\g> 0$ be the coupling constant,  $\ka=1$ be the  ultraviolet cut-off\footnote{We set $\ka=1$ to simplify
the proofs of Proposition~\ref{preliminaries-on-spectrum} and Theorem~\ref{main-theorem-spectral}, given in the companion paper \cite{DP12}. 
In the present paper we will write $\ka$ explicitly.}
 and let $1/2\geq \alf\geq 0$ be a parameter which controls the infrared behavior of the system. Given these 
parameters, we define the form-factor
\beqa
\vv(k):=\g\fr{  \chi_{\ka}(k) |k|^{\alf} }{(2|k|)^\half}, \label{form-factor-formula}
\eeqa
where $\chi_{\ka}\in C_0^{\infty}(\real^3)$ is rotationally invariant, non-increasing in the radial direction, supported in $\mcB_\ka$ and equal to one on $\mcB_{(1-\eps_0)\ka}$, for some fixed $0<\eps_0<1$. (We denote by $\mcB_r$ the open ball of radius $r$ centered at zero).

The interaction Hamiltonian, defined  as a symmetric  operator on $\mcC$, is given by the following formula 
\beqa
H_{\I}:=\int d^3p d^3k\,\vv(k)\nn^*(p+k)\bb(k)\nn(p)+\hc \label{interaction-Hamiltonian}
\eeqa
For future reference we denote by $H_{\I}^\ain$ be the first term on the r.h.s. of (\ref{interaction-Hamiltonian}) and
set $H_{\I}^\cin=(H_{\I}^\ain)^*$.

As indicated in \cite{Fr73,Fr74.1}, the full Hamiltonian $H=H_{\free}+H_{\I}$  can be defined as a self-adjoint 
operator on a dense domain in $\hil$. For the reader's convenience we outline briefly this construction: First,
we note that both $H_{\free}$ and $H_{\I}$ preserve the number of electrons. Let us therefore define
$\hil^{(n)}:=\Ga^{(n)}(\mfh_{\el})\otimes \Ga(\mfh_{\pho})$, where $\Ga^{(n)}(\mfh_{\el})$
is the $n$-particle subspace of $\Ga(\mfh_{\el})$ and let $H_{\free}^{(n)}$ and $H_{\I}^{(n)}$ be the restrictions
of the respective operators to $\hil^{(n)}$, defined of $\mcC^{(n)}:=\mcC\cap \hil^{(n)}$. As shown in Lemma~\ref{self-adjointness},
using the Kato-Rellich theorem, each $H^{(n)}=H_{\free}^{(n)}+H_{\I}^{(n)}$ can be defined as a bounded from below, 
self-adjoint operator on the domain of $H^{(n)}_{\free}$, which is essentially self-adjoint on $\mcC^{(n)}$. Then we can define
\beqa
H:=\bigoplus_{n\in \nat_0} H^{(n)}
\eeqa 
as an operator on $\mcC$.  Since $H^{(n)}\pm i$ have  dense ranges  on $\mcC^{(n)}$,  $H\pm i$ has a dense range on $\mcC$, 
thus it is essentially self-adjoint on this domain. We stress that the above construction is valid both in the infrared-regular case ($1/2\geq \alf>0$)
and in the infrared singular sitiation ($\alf=0$).

On $\mcC$ we have the following formula for $H$
\beqa
H=\int d^3p\, \Om(p)\nn^*(p)\nn(p)&+&\int d^3k\, \om(k) a^*(k)a(k)\non\\
&+&(\int d^3p d^3k\,\vv(k)\nn^*(p+k)\bb(k)\nn(p)+\hc). \label{full-Nelson-model}
\eeqa
It  reduces to a more familiar expression on $\mcC^{(n)}$
\beqa
H^{(n)}=\sum_{i=1}^n\fr{(i\nabla_{x_i})^2}{2}+\int d^3k\, \om(k) a^*(k)a(k)+\sum_{i=1}^n\int d^3k\, \vv(k)\, ( e^{ikx_i}a(k)+e^{-ikx_i}a^*(k) ),
\label{explicit-Hamiltonian}
\eeqa
where $x_i$ are the position operators of the electrons. Finally, we introduce the electron and photon momentum operators
\beqa
P_{\el}^i:=\int d^3p\, p^i\nn^*(p)\nn(p),\quad P_{\pho}^i:=\int d^3k\, k^i a^*(k)a(k),\quad i\in\{1,2,3\},
\eeqa
 which are essentially self-adjoint  on $\mcC$. We  recall that $H$ is translationally invariant, that is it commutes with the
total momentum operators $P^i$, given by
\beqa
P^i:=P_{\el}^i+P_{\pho}^i,\quad  i\in\{1,2,3\},
\eeqa
which are also essentially self-adjoint on $\mcC$.


\subsection{Spectral theory}\label{spectral-sub}

In this section we collect some necessary information about the spectrum of the single-electron Hamiltonian $H^{(1)}$.
We recall that the analysis of the spectrum of $H_{P}^{(1)}$ was initiated in \cite{Fr73,Fr74.1} and advanced in \cite{Pi03}   
with the help of the iterative analytic perturbation theory. Further developments along these lines, which we describe in
detail below, can be found in \cite{DP12}. Interesting results on the spectrum of the  Nelson model with a slightly different
form factor were also obtained in \cite{AH12} by different methods.

 We recall that due to the translational invariance $H^{(1)}$ can be decomposed into a direct integral of the
fiber Hamiltonians $H_{P}^{(1)}$ as follows
\beqa
H^{(1)}=\Pi^* \int^{\oplus} d^3P\, H_{P}^{(1)} \,\, \Pi.
\eeqa
Here $\Pi:\hil^{(1)}\to L^2(\real^3;\Ga(\mfh_{\fib}))$ is a unitary map given by $\Pi:=F e^{iP_{\pho}x}$, where $x^i$ are the electron's position operators 
and $F$ is the Fourier transform in the electron's variables. The Hamiltonians $H_{P}^{(1)}$  are self-adjoint operators on the Fock space $\Ga(\mfh_{\fib})$, where $\mfh_{\fib}=L^2(\real^3,d^3k)$ is the single-photon space in the fiber picture. Denoting by $b^*(k)$ and $b(k)$ the creation
and annihilation operators on $\Ga(\mfh_{\fib})$ one easily obtains from (\ref{explicit-Hamiltonian}) that
\beqa
H_{P}^{(1)}=\h(P-P_{\pho})^2+H_{\pho}+\int d^3k\, \vv(k)\, ( b(k)+b^*(k) ).
\eeqa
As a tool to study  the spectrum of this Hamiltonian, we introduce auxiliary fiber Hamiltonians with infrared cut-offs
\beqa
H_{P,\si}^{(1)}=\h(P-P_{\pho})^2+H_{\pho}+\int d^3k\, \vv^{\si}(k)\, ( b(k)+b^*(k) ). \label{infrared-cut-off-Hamiltonian}
\eeqa  
The form-factor $\vv^{\si}$  is defined as follows
\beqa
\vv^{\si}(k):=\g\fr{ \chi_{[\si,\ka)}(k)    |k|^{\alf} }{(2|k|)^\half}, \label{IR-cut-off-propagator}
\eeqa
where $0<\si\leq \ka$, $\chi_{[\si,\ka)}(k):=\mathbf{1}_{\mcB'_{\si}   }(k)\chi_{\ka}(k)$, $\mcB_\si'$ is the complement of the ball of radius $\si$
and $\mathbf{1}_{\De}$ is the characteristic function of a set $\De$. By replacing $\vv$ with $\vv^{\si}$ in (\ref{interaction-Hamiltonian}),
we obtain the interaction Hamiltonian $H_{\I,\si}$ and the corresponding full Hamiltonian $H_{\si}$ with an infrared cut-off. The restriction
of $H_{\si}$ to the single-electron subspace, denoted  $H_{\si}^{(1)}$, has a fiber decomposition into
the Hamiltonians $H_{P,\si}^{(1)}$.

Since we are particularly interested in the bottom of the spectrum of $H_P^{(1)}$ and $H_{P,\si}^{(1)}$, let us define
\beqa
 E_{P}:=\inf \si(H_{P}^{(1)}), \quad  E_{P,\si}:=\inf \si(H_{P,\si}^{(1)}).
\eeqa
As the model is non-relativistic, we restrict attention to small values of the total momentum $P$ at which the electron 
moves slower than the photons.   More precisely, we consider $P$ from the set
\beqa
\PS:=\{\, P\in \real^3\,|\, |P|< P_{\maxi}\,\}
\eeqa
for some $P_{\maxi}>0$. Since we  work in the weak coupling regime, we  fix some sufficiently small $\g_0>0$, 
and restrict attention to $\g\in (0, \g_0]$. The parameters $P_{\maxi}$ and $\g_0$ are specified in 
Proposition~\ref{preliminaries-on-spectrum} and Theorem~\ref{main-theorem-spectral} below. $P_{\maxi}$
remains fixed in the course of our analysis. The maximal coupling constant $\g_0$ is readjusted only in the
last step of our investigation -- in Theorem~\ref{convergence-main-theorem} -- to a new value which is denoted $\g_0'$.

In the following proposition we collect the results concerning $E_P, E_{P,\si}$ 
which will be needed in the present investigation. Many of these properties are known, but several are new, as we explain below.
\bep\label{preliminaries-on-spectrum} Fix $0\leq \alf \leq 1/2$ and let $P_{\maxi}=1/6$. Then there exists $\g_0>0$ s.t. for all $P\in S:=\mcB_{P_{\maxi}}$,
$\g\in (0, \g_0]$  there holds:
\begin{enumerate}[label = \textup{(\alph*)}, ref =\textup{(\alph*)},leftmargin=*]
\item\label{energy-part} $S\ni P\mapsto E_P$ is twice continuously differentiable and strictly convex.
$S\ni P\mapsto E_{P,\si}$ is analytic and strictly convex, uniformly in $\si\in (0,\ka]$.  Moreover, 
\beqa
& &|E_{P}-E_{P,\si}|\leq c\si, \label{energy-convergence-bound}\\
& &|\pa_{P}^{\be_1}E_{P,\si}|\leq c,\quad  |\pa_{P}^{\be_2}E_{P,\si}|\leq c, \quad  |\pa_{P}^{\be_3}E_{P,\si}|\leq c/\si^{\de_{\la_0}} 
\label{velocity-boundedness}
\eeqa
for multiindices $\be_j$ s.t. $|\be_j|=j$, $j\in\{1,2,3\}$.

\item\label{cut-off-part} For $\si>0$, $E_{P,\si}$ is a simple eigenvalue corresponding to a normalized eigenvector $\psi_{P,\si}$, whose
phase is specified in \cite{DP12}. 
There holds
\beqa
\|\pa^{\be}_P\psi_{P,\si}\|\leq  c /\si^{\de_{\g_0}} \label{state-bound}
\eeqa
for  multiindices $\be$ s.t. $0<|\be|\leq 2$.

\item\label{eigenvectors-convergence} For $\alf>0$, $E_P$ is a simple eigenvalue corresponding to a normalized eigenvector $\psi_{P}$.
Moreover, for a suitable choice of the phase of $\psi_P$,
\beqa
\|\psi_P-\psi_{P,\si}\|\leq c\si^{\alf}.
\eeqa

\end{enumerate}
The constant $c$ above is independent of $\si$, $P$, $\g$, $\alf$ within the assumed restrictions. Clearly, all statements above
remain true after replacing $\g_0$ by some $\ti\g_0\in (0,\g_0]$. The resulting function
  $\ti\g_0\mapsto \de_{\ti\g_0}$ can be chosen positive and s.t. $\lim_{\ti\g_0\to 0} \de_{\ti\g_0}=0$.
\eep
In the light of Proposition~\ref{preliminaries-on-spectrum} we can define the subspace of renormalized 
single-electron states
\beqa
\hil_{1,\si}:=\{\Pi^*\int^{\oplus}d^3P\, h(P)\psi_{P,\si}\,|\, h\in L^2(\real^3,d^3P), \supp\, h\subset \PS\,\}. \label{single-electron-space}
\eeqa 
In the case of $1/2\geq \alf>0$ we also set $\hil_{1}:=\hil_{1,\si=0}$.

Large part of Proposition~\ref{preliminaries-on-spectrum} has already been established  in the
Nelson model or in similar models: The fact that $S\ni P\mapsto E_P$ is twice continuously differentiable and convex has been
shown  in non-relativistic and semi-relativistic QED in \cite{FP10,KM12, BCFS07} and in the
Nelson model with a slightly different form-factor in  \cite{AH12}. The present case
is covered by our analysis in \cite{DP12}. The bound (\ref{energy-convergence-bound})
can be extracted from \cite{Pi03}. First statement in \ref{cut-off-part}  has been established already in \cite{Fr74.1}.
Part \ref{eigenvectors-convergence} is implicit in \cite{Pi03} and is shown explicitly in \cite{DP12}. 
The bound on the third derivative of $E_{P,\si}$ in (\ref{velocity-boundedness}) and on the first and second derivative
of $\psi_{P,\si}$ in (\ref{state-bound}) are new and are proven in \cite{DP12}.

It turns out that the properties stated in Proposition~\ref{preliminaries-on-spectrum} are not quite enough 
for our purposes. Scattering theory for several electrons requires much more detailed information about
the electron's localization in space than scattering of one electron and photons. This information is
contained in  regularity properties of the momentum wave functions of the  vectors $\psi_{P,\si}$.
Let us express $\psi_{P,\si}$ in terms of its $m$-particle components in the Fock space: 
\beqa
\psi_{P,\si}=\{f^m_{P,\si}\}_{m\in \nat_0},
\eeqa
where $f^m_{P,\si}\in L^2_{\sym}(\real^{3m}, d^{3m}k)$, i.e., each $f^m_{P,\si}$   is a square-integrable function symmetric in $m$ variables from $\real^3$. Let us introduce the following auxiliary functions:
\beqa
 g^m_{\si}(k_1,\ldots, k_m):=\prod_{i=1}^m\fr{ c\g  \chi_{[\si,\kas)}(k_i) |k_i|^{\alf} }{|k_i|^{3/2}},\quad \kas:=(1-\eps_0)^{-1}\ka, \quad
0<\eps_0<1, \label{def-g-ks}
\eeqa
where $\tic$ is some positive constant independent of $m,\si, P$ and $\g$ within the restrictions specified above. 
Finally, we introduce the notation
\beqa
\mcA_{r_1,r_2}:=\{\, k\in \real^3\,|\, r_1<|k|<r_2 \,\}, \label{A-set}
\eeqa
where $0\leq r_1< r_2$. 
Now we are ready to state the required properties of the functions $f^m_{P,\si}$:
\bet \label{main-theorem-spectral} 
Fix $0\leq \alf \leq 1/2$ and set $P_{\maxi}=1/6$. Then there exists  $\lambda_0>0$ s.t. for all $P \in S=\mathcal{B}_{P_{\maxi}}$,
$\lambda\in (0, \lambda_0]$  there holds:
\begin{enumerate}[label = \textup{(\alph*)}, ref =\textup{(\alph*)},leftmargin=*]

\item \label{f-m-support} Let  $\{f^m_{P,\sigma}\}_{m\in\nat_0}$ be the $m$-particle components of $\psi_{P,\sigma}$
and let $\overline{\mathcal{A}}_{\sigma,\kappa}^{\times m}$ be defined by (\ref{A-set}). Then,
  for any $P\in S$, the function $f^m_{P,\sigma}$ is supported in $\overline{\mathcal{A}}_{\sigma,\kappa}^{\times m}$.

\item \label{f-m-smoothness} The  function
\begin{equation}
S\times \mathcal{A}_{\sigma,\infty}^{\times m}\ni (P; k_1,\dots, k_m) \mapsto f^m_{P,\sigma}(k_1,\dots, k_m) \label{momentum-wave-functions}
\end{equation}
is twice continuously differentiable and extends by continuity, together with its derivatives, to the set $S\times \overline{\mathcal{A}}_{\sigma,\infty}^{\times m}$. 

\item\label{derivatives-bounds} For any multiindex $\beta$, $0\leq |\beta|\leq 2$, the function (\ref{momentum-wave-functions}) satisfies
\begin{eqnarray}
|\partial^{\beta}_{k_l}f^m_{P,\sigma}(k_1,\dots, k_m)|&\leq& \frac{1}{\sqrt{m!}} |k_l|^{-|\beta|}g^m_{\sigma}(k_1,\dots, k_m), \label{simple-spectral-bound}\\
|\partial^{\beta}_{P}f^m_{P,\sigma}(k_1,\dots, k_m)|&\leq&  \frac{1}{\sqrt{m!}} \big( \frac{1}{\sigma^{\delta_{\delta_0}}}\big)^{| \beta |}  g^m_{\sigma}(k_1,\dots, k_m), \label{infrared-spectral-bound}\\
|\partial_{P^{i'}}\partial_{k_l^i}f^m_{P,\sigma}(k_1,\dots, k_m)|& \leq &  \frac{1}{\sqrt{m!}}\frac{1}{\sigma^{\delta_{\delta_0}}}|k_l|^{-1}g^m_{\sigma}(k_1,\dots, k_m), \label{mixed-spectral-bound}
\end{eqnarray}
where the function $\tilde{\lambda}_0 \mapsto \lambda_{\tilde{\delta_0}}$ has the properties specified in Proposition~\ref{preliminaries-on-spectrum}.
\end{enumerate}
\eet
Parts \ref{f-m-support}, \ref{f-m-smoothness} of Theorem~\ref{main-theorem-spectral} and   estimate~(\ref{simple-spectral-bound}) in \ref{derivatives-bounds}  can be extracted from \cite{Fr73,Fr74.1,Fr73.1} or proven using the methods from these papers. The key new input are the bounds (\ref{infrared-spectral-bound}), (\ref{mixed-spectral-bound})
on the first and second derivative w.r.t. $P$ with their mild dependence on the infrared cut-off $\si$.  These bounds
require major refinements of the iterative multiscale analysis from \cite{Pi03} and they constitute the main technical result
of the companion paper \cite{DP12}.

\subsection{Scattering theory}\label{scatter-sub}

In this section we outline the construction of  two-electron scattering states  along the lines of Haag-Ruelle
scattering theory \cite{Ha58, Ru62}, following some ideas from \cite{Fr73.1, Al73}. Since we are interested here in the
infrared-regular situation, we set $1/2\geq \alf>0$. Such $\alf$ will be kept fixed in the remaining part of the paper.
(The infrared-singular case $\alf=0$ is much
more involved due to the infraparticle problem and will be studied elsewhere). 

Let $\B(\PS)$ be the class of twice continuously differentiable functions with compact support contained  in $\PS$.
Following \cite{Fr73.1, Al73}, for any $h\in \B(\PS)$ we define the renormalized creation operator
\beq
\nr_{\si}^*(h):=\sum_{m=0}^{\infty}\fr{1}{\sqrt{m!}}\int d^3p\,d^{3m}k\,h(p) f^{m}_{p,\si}(k)\bb^*(k)^m \, \nn^*(p- \unk).
\label{renormalized}
\eeq 
In this expression we use the following short-hand notation, which will appear frequently below:
\beqa
& &f^m_{p,\si}(k):=f^m_{p,\si}(k_1,\ldots, k_m),\\
& &\bb^*(k)^m:=\bb^*(k_1)\ldots\bb^*(k_m), \\
& &\un k:=k_1+\cdots +k_m.
\eeqa
It is shown in Lemma~\ref{ren-creation-operator} that $\nr_{\si}^*(h)$ and $\nr_{\si}^*(h_1)\nr_{\si}^*(h_2)$, for $h_1, h_2\in \B(\PS)$,
 are well defined operators  on $\mcC$. (Since $\nr_{\si}(h):=(\nr_{\si}^*(h))^*$ is obviously well defined on $\mcC$, we obtain that
$\nr_{\si}^*(h)$ is closable).

Now let $\vac_{\el}$ and $\vac_{\pho}$ be the vacuum vectors of $\Ga(\mfh_{\el})$ and $\Ga(\mfh_{\pho})$, respectively. Then
$\vac:=\vac_{\el}\otimes \vac_{\pho}$ is the physical vacuum in $\hil$. As shown in Lemma~\ref{fiber-ground-states-lemma} below,
\beqa
\psi_{h,\si}:=\nr_{\si}^*(h)\vac=\Pi^*\int^{\oplus}d^3P\,h(P)\psi_{P,\si}, 
\eeqa
that is $\psi_{h,\si}$ is an element of the subspace of renormalized single-electron states $\hil_{1,\si}$, 
defined in~(\ref{single-electron-space}). Since we assumed that $1/2\geq \alf>0$, we obtain from 
Proposition~\ref{preliminaries-on-spectrum} \ref{eigenvectors-convergence} that there exists the limit
\beqa
\psi_{h}:=\slim_{\si\to 0}\nr_{\si}^*(h)\vac=\Pi^*\int^{\oplus}d^3P\,h(P)\psi_{P}. \label{single-electron-states}
\eeqa
Clearly, $\psi_{h}$ belongs to the renormalized single-electron space $\hil_{1}$ of the Hamiltonian $H$.

Let us now proceed to the construction of two-electron scattering states. We fix some parameter $\ga_0>4$,
which will be fixed in our investigation, choose some $\ga\in (4,\ga_0]$ and  introduce a time-dependent cut-off 
\beqa
\si_t:=\siz / t^{\gamma} \label{time-dependent-cut-off}
\eeqa
for $t\geq \max\{1,\siz\}$.
Next, we choose $h_1, h_2\in \B(\PS)$, with disjoint supports and set 
\beqa
h_{i,t}(p):=e^{-iE_p t}h_i(p),\quad i\in\{1,2\}.
\eeqa
Now we are ready to  define the two-electron scattering states approximants:
\beqa
\Psi_{t,h_1,h_2}:=e^{iHt}\nr_{\si_t}^*(h_{1,t})\nr_{\si_t}^*(h_{2,t})\vac. \label{scattering-states-approximants}
\eeqa
We will show that the  limit of $\Psi_{t,h_1,h_2}$ exists as $t\to \infty$ and can be interpreted 
as a physical state describing two independent excitations. This is the content of our main result
concerning scattering theory, stated below. 
\bet\label{main-result-scattering} Fix $1/2\geq \alf>0$, $\ga_0>4$. Let $\la\in (0,\la_0']$, where $\la_0'>0$ is sufficiently small.   
Then, for $h_1, h_2\in  \B(\PS)$ with disjoint supports, the following statements hold: 
\begin{enumerate}[label = \textup{(\alph*)}, ref =\textup{(\alph*)},leftmargin=*]

\item\label{scattering-states-existence}  Let $\si_t=\ka/t^{\ga}$, where $\ga\in (4,\ga_0]$. Then there exists the limit
\beqa
\Psi^+_{h_1,h_2}:=\slim_{t\to\infty} e^{iHt}\nr_{\si_t}^*(h_{1,t})\nr_{\si_t}^*(h_{2,t})\vac \label{scattering-state-convergence}
\eeqa 
and it is called the two-electron scattering state. It is independent of the parameter $\ga$ within the above restrictions.

\item\label{scalar-product} Let  $\Psi^+_{h_1,h_2}$, $\Psi^+_{h_1',h_2'}$ be two scattering states. Their scalar product has the form
\beqa
\lan \Psi^+_{h_1,h_2}, \Psi^+_{h_1',h_2'}\ran=\lan \psi_{h_1},\psi_{h_1'}\ran \lan \psi_{h_2},\psi_{h_2'}\ran+
\lan \psi_{h_1},\psi_{h_2'}\ran \lan \psi_{h_2},\psi_{h_1'}\ran.
\eeqa 

\end{enumerate}

\eet
\proof Part \ref{scattering-states-existence} follows from Theorem~\ref{convergence-main-theorem}.
Assumptions~(\ref{tensor-product-structure}), (\ref{tensor-product-rest-term}) of this theorem are verified in
Proposition~\ref{scalar-product-proposition}. Assumption~\ref{cook-method} follows from Propositions~\ref{Cook-method-proposition},
\ref{double-commutator-proposition}, \ref{check-contribution} and Corollary~\ref{h-sigma-part}. Part \ref{scalar-product} of the
theorem follows from Proposition~\ref{scalar-product-proposition} and Proposition~\ref{preliminaries-on-spectrum} \ref{eigenvectors-convergence}.\qed\\ 
We remark that an essential ingredient of the proof of convergence in (\ref{scattering-state-convergence})
is the disjointness of  velocity supports of the functions $h_1$, $h_2$, defined as
\beqa
V(h_i):=\{\,\nabla E_{p}\, |\, p\in \supp\,h_i \,\},\quad i\in\{1,2\}. \label{velocity-support-one}
\eeqa
This property follows from the assumed disjointness of the supports of $h_1, h_2$ and the invertibility of the
relation $\PS\ni p\mapsto E_p$, guaranteed by Proposition~\ref{preliminaries-on-spectrum} \ref{energy-part}.

We note that the vector $\psi_h$, given by (\ref{single-electron-states}), is determined uniquely by the function $h\in \B(\PS)$. 
Moreover, vectors of the form $\psi_{h_1}\otimes_s\psi_{h_2}$, where $h_1, h_2\in \B(\PS)$ have disjoint supports, and
$\otimes_s$ is the symmetric tensor product, span a dense subspace in $\hil_{1}\otimes_s\hil_{1}$. Thus we can define the wave operator 
$W^+: \hil_{1}\otimes_s\hil_{1}\to \hil$  as follows 
\beqa
W^+ (\psi_{h_1}\otimes_s \psi_{h_2}):=\Psi^+_{h_1,h_2}.
\eeqa
By Theorem~\ref{main-result-scattering} \ref{scalar-product}, this map is an isometry.

\vspace{0.5cm}

\noindent \bf Standing assumptions and conventions\rm: 
\begin{enumerate}

\item The  parameters $P_{\maxi}=1/6$ and  $1/2\geq \alf>0$ are kept fixed in the remaining part of the paper. 

\item The maximal coupling constant $\g_0>0$ is as specified in Proposition~\ref{preliminaries-on-spectrum} and Theorem~\ref{main-theorem-spectral}
for the values of $P_{\maxi}$ and $1/2\geq \alf>0$ fixed above. It remains unchanged in Sections~\ref{vacuum-expectation-values} and \ref{non-stationary-phase}.
It is readjusted to a possibly smaller value $\g_0'>0$ in the last step of the analysis in Theorem~\ref{convergence-main-theorem}.
$\g_0'$ may depend of $\ga_0$ but not on $\ga\in (4,\ga_0]$.

\item $\ti\g_0\mapsto \de_{\ti\g_0}$, $\ti\g_0\mapsto \de'_{\ti\g_0}$, will denote positive functions of $\ti\g_0\in (0,\la_0]$, which may differ from line to line,
 and have the property
\beqa
\lim_{\ti\g_0\to 0} \de_{\ti\g_0}=0,  \quad \lim_{\ti\g_0\to 0} \de'_{\ti\g_0}=0.
\eeqa
Such functions control the infrared behaviour of our estimates listed in the statement of Theorem~\ref{convergence-main-theorem}. By reducing  the coupling constant from $\g_0$ to $\g_0'$ we make this behaviour sufficiently mild.

\item We denote by $\ga\in (4,\ga_0]$ the parameter which controls the time dependence of the (fast) infrared cut-off i.e., $\si_t=\siz/t^{\ga}$.
The parameter $\ga_0$ is kept fixed in the remaining part of the paper. This parameter appears also in the definition of the
slow infrared cut-off $\si_{\s}=\ka(\si/\ka)^{1/(8\ga_0)}$ in the proofs of  Lemmas~\ref{smooth-bounds} and \ref{second-contribution-decay}.

\item We will denote by $c,c',c''$ numerical constants which may depend on $\PS$, $\g_0$, $\eps_0$, $\ka$, $\alf$,  $\tiga$ and functions $h_1,h_2$ 
but not on $\si$, $t$ or the electron and photon momenta. The values of these constants may change from line to line.

\item We will denote by $(p,q)\mapsto D(p,q)$, $(p,q)\mapsto D'(p,q)$ smooth, compactly supported functions on $\real^3\times\real^3$,
which may depend of $\PS$,$\g_0$, $\eps_0$,  $\ka$, $\alf$ but not on $\si$, $t$ or  photon momenta.

\item We will denote by $k=(k_1,\ldots, k_m)\in \real^{3m}$ a collection of photon variables.
A lower or upper index $m$ of a function indicates that it is a symmetric  function of $(k_1,\ldots, k_m)$.  
For example:
\beqa
f^m(k):=f^{m}(k_1,\ldots, k_m).
\eeqa  
Similarly, we set $a^*(k)^m:=a^*(k_1)\ldots a^*(k_m)$. We note that the order in which the components of $k$ are
listed is irrelevant, since they enter always into symmetric expressions.

\item We separate the electron variable $p\in\real^3$ and the photon variables $k\in \real^{3m}$ by a semicolon.
For example:
\beqa
G_m(p; k):=G_m(p; k_1,\ldots, k_m). \label{notation-G-m}
\eeqa

\item Two collections of electron and photon variables $p\in\real^3, k\in \real^{3m}$ and $q\in\real^3, r\in \real^{3n}$ are
separated by a bar. For example
\beqa
F_{m,n}(p;k \ba q;r):=F_{m,n}(p;k_1,\ldots, k_m \ba q; r_1,\ldots, r_n).
\eeqa

\item Given $k=(k_1,\ldots, k_m)$ we write $\un k:= k_1+\cdots +k_m$.

\end{enumerate}

\section{Main ingredients of the proof}\label{main-steps}
\setcounter{equation}{0}

Theorem~\ref{convergence-main-theorem} below gives the existence of scattering states based on some assumptions 
proven in the later part of this paper: Property~(\ref{tensor-product-structure}) follows from Proposition~\ref{scalar-product-proposition}.
To establish property~(\ref{cook-method}) we derive a formula for time-derivatives of scattering state approximants 
 in  Proposition~\ref{Cook-method-proposition} in the later part of this section. The section concludes with Lemma~\ref{fiber-ground-states-lemma}
which gives a direct integral representation of the vectors $\psi_{h,\si}=\nr_{\si}^*(h)\vac$. We used this result already in (\ref{renormalized})
above.

In our proof of convergence of the scattering state approximants~(\ref{scattering-states-approximants}) we will vary time $t$ and the 
infrared cut-off $\si$ independently. Let us therefore  
 fix $h_1,h_2\in \B(\PS)$ with disjoint supports and introduce an auxiliary two-parameter sequence
\beqa
\Psi_{t,\si}:=e^{iHt}\nr_{\si}^*(h_{1,t})\nr_{\si}^*(h_{2,t})\vac. \label{two-par-approx}
\eeqa
Now we are ready to state and prove the main result of this section.
\bet\label{convergence-main-theorem} Let $h_1,h_2\in \B(\PS)$ have disjoint supports, let $\Psi_{t,\si}$ be given by (\ref{two-par-approx}). 
Suppose that for $\la\in (0,\la_0]$, infrared cut-offs $\si, \si'$ s.t. $\si\leq \si'\leq \siz$ and  $t\geq \max\{1,\siz\}$ 
\beqa
\lan\Psi_{t,\sip},\Psi_{t,\si}\ran&=&\lan   \psi_{h_1,\sip},   \psi_{h_1,\si}\ran 
\lan  \psi_{h_2,\sip}, \psi_{h_2,\si}\ran+R(t,\si,\si'),  \label{tensor-product-structure}\\
|R(t,\si,\si')|&\leq& c\fr{1}{\si^{\de_{\g_0} }} \bigg( \fr{1}{t}\fr{1}{\si^{1/(8\tiga)} }+ (\si')^{2\alf}  \bigg), \label{tensor-product-rest-term} \\   
\|\pa_t\Psi_{t,\si}\|&\leq&\fr{c}{\si^{\de_{\g_0}}  } \bigg( \fr{\si^{\alf/(4\tiga)  } }{t}+\si t +\fr{1}{t^2\si^{1/(4\tiga)}} \bigg)+ c\si^{1-\de_{\g_0}}
+c\si^{1/2-\de_{\la_0}}\bigg( 1+ \fr{1}{t}\fr{1}{\si^{1/(8\tiga)} }  \bigg). \,\,\,\,\,\,\, \,\,\label{cook-method} 
\eeqa
Then one can choose $\g_0'\in (0,\g_0]$ s.t. for $\g \in (0,\g_0']$  there exists the limit
\beqa
\Psi^+_{h_1,h_2}=\slim_{t\to\infty}\Psi_{t,\si_t},
\eeqa
where $\si_t=\siz/t^\ga$. This limit is independent of the choice of  $\ga\in (4,\ga_0]$.
\eet
\proof   We assume that $t_2\geq t_1\geq 1$ are sufficiently large so that $\si_{t_2}\leq \si_{t_1}\leq 1$. We write
\beqa
\|\Psi_{t_2,\si_{t_2}}-\Psi_{t_1,\si_{t_1}}\|\leq   \|\Psi_{t_2,\si_{t_2}}-\Psi_{t_1,\si_{t_2}}\|+ \| \Psi_{t_1,\si_{t_2}}-  \Psi_{t_1,\si_{t_1}}\|.
\label{Cauchy}
\eeqa
Concerning the first term on the r.h.s. of (\ref{Cauchy}), we note that the bound in (\ref{cook-method}) implies
\beqa
\|\pa_t \Psi_{t,\si_{t_2}}\|\leq\fr{c}{\si_{t_2}^{\de_{\g_0}}  } \bigg( \fr{\si_{t_2}^{\eps} }{t}+\fr{1}{t^2\si_{t_2}^{1/(4\ga_0)}} \bigg)+ c\si_{t_2}^{1/2-\de_{\la_0}}t
\eeqa
for some $\eps>0$, depending on $\ga_0$, but independent of $\g_0$. Now we estimate
\beqa
\|\Psi_{t_2,\si_{t_2}}-\Psi_{t_1,\si_{t_2}}\|&\leq& \int^{t_2}_{t_1}dt\, \|\pa_t \Psi_{t,\si_{t_2}}\|
\leq   c\si_{t_2}^{\eps-\de_{\g_0}} \log(t_2/t_1)   +c\fr{1}{t_1}\fr{1}{\si_{t_2}^{ \de_{\g_0}+1/(4\ga_0) }  }  +c\si_{t_2}^{1/2-\de_{\la_0}}t_2^2\non\\
&\leq& c\si_{t_2}^{\eps'}    +c\fr{1}{t_1}\fr{1}{\si_{t_2}^{ 1/(3\ga_0) }  }  +c\si_{t_2}^{1/2-\de_{\la_0}}t_2^2
\leq c\fr{1}{t_2^{\eps''}}+c\fr{t_2^{1/3}}{t_1}\non\\
&\leq &c\bigg(\fr{t_2^{1/3}}{t_1}+\bigg(\fr{t_2^{1/3}}{t_1}\bigg)^{\eps''}\bigg), 
\label{Cook-method-argument}
\eeqa
where in the third step we made use of the fact that $t_2^{-\al} \log(t_2/t_1)$ is uniformly bounded in 
$t_2\geq t_1\geq 1$ for any $\al>0$ and chose $\la_0$ sufficiently small (depending on $\ga_0$) to ensure that $\eps'>0$
and that $\de_{\g_0}+1/(4\ga_0) \leq 1/(3\ga_0) $. In the fourth step we chose $\g_0$ again sufficiently small  and exploited the fact that $\ga>4$ to ensure that $1>\eps''>0$.

As for the second term on the r.h.s. of (\ref{Cauchy}), we get
\beqa
& &\| \Psi_{t_1,\si_{t_2}}-  \Psi_{t_1,\si_{t_1}}\|^2\non\\
&=&\lan \Psi_{t_1,\si_{t_2}},\Psi_{t_1,\si_{t_2}} \ran
+ \lan \Psi_{t_1,\si_{t_1}}, \Psi_{t_1,\si_{t_1}}\ran
-2\mathrm{Re}\lan \Psi_{t_1,\si_{t_2}}, \Psi_{t_1,\si_{t_1}} \ran \non\\
&=&\lan  \psi_{h_1,\si_{t_2}},  \psi_{h_1,\si_{t_2}}\ran 
\lan \psi_{h_2,\si_{t_2}}, \psi_{h_2,\si_{t_2}}\ran+R(t_{1},\si_{t_2},\si_{t_2})\non\\
& &+ \lan  \psi_{h_1,\si_{t_1}},  \psi_{h_1,\si_{t_1}}\ran 
\lan \psi_{h_2,\si_{t_1}}, \psi_{h_2,\si_{t_1}}\ran+R(t_{1},\si_{t_1},\si_{t_1})\non\\
& &-\lan  \psi_{h_1,\si_{t_2}},   \psi_{h_1,\si_{t_1}}\ran 
\lan  \psi_{h_2,\si_{t_2}}, \psi_{h_2,\si_{t_1}}\ran+\ov{R(t_1,\si_{t_2},\si_{t_1})}\non\\
& &-\lan   \psi_{h_1,\si_{t_1}},   \psi_{h_1,\si_{t_2}}\ran 
\lan  \psi_{h_2,\si_{t_1}},    \psi_{h_2,\si_{t_2}}\ran+R(t_1,\si_{t_2},\si_{t_1}).
\eeqa
Thus making use of Proposition~\ref{preliminaries-on-spectrum} \ref{eigenvectors-convergence} and (\ref{tensor-product-rest-term}), we obtain
\beqa
\| \Psi_{t_1,\si_{t_2}}-  \Psi_{t_1,\si_{t_1}}\|^2&\leq&c\si_{t_1}^{\alf}+|R(t_{1},\si_{t_2},\si_{t_2})|+|R(t_{1},\si_{t_1},\si_{t_1})|
+2|R(t_1,\si_{t_2},\si_{t_1})|\non\\
&\leq&c\si_{t_1}^{\alf}+c\fr{1}{\si_{t_2}^{\de_{\g_0} }} \bigg( \fr{1}{t_1}\fr{1}{\si_{t_2}^{1/(8\ga_0)} }+ (\si_{t_2})^{2\alf}  \bigg)\non\\
& &+c\fr{1}{\si_{t_1}^{\de_{\g_0} }} \bigg( \fr{1}{t_1}\fr{1}{\si_{t_1}^{1/(8\ga_0)} }+ (\si_{t_1})^{2\alf}  \bigg)\non\\
& &+c\fr{1}{\si_{t_2}^{\de_{\g_0} }} \bigg( \fr{1}{t_1}\fr{1}{\si_{t_2}^{1/(8\ga_0)} }+ (\si_{t_1})^{2\alf}  \bigg).
\eeqa
Thus, by choosing $\g_0$ sufficiently small, we can find $0<\de_1,\de_2,\de<1$ s.t.
\beqa
\| \Psi_{t_1,\si_{t_2}}-  \Psi_{t_1,\si_{t_1}}\|^2&\leq& 
c\bigg(\fr{1}{t_1^{\de_1}}+\fr{t_2^{1/3}}{t_1}+\fr{1}{t_2^{\de_2}}+\fr{t_2^{\de_{\la_0}}}{t_1^{\de_1}}\bigg)\non\\
&\leq& c\bigg(\fr{t_2^{\de_{\la_0}}}{t_1^{\de}}+ \fr{t_2^{1/3}}{t_1}\bigg)\leq c\bigg(\bigg(\fr{t_2^{1/3} }{t_1}\bigg)^{\de} + \fr{t_2^{1/3} }{t_1}\bigg),  
\eeqa
where in the second step we used that $t_2\geq t_1$. Consequently
\beqa
\| \Psi_{t_1,\si_{t_2}}-  \Psi_{t_1,\si_{t_1}}\|\leq c\bigg(\bigg(\fr{t_2^{1/3} }{t_1}\bigg)^{\de/2} +\bigg( \fr{t_2^{1/3} }{t_1}\bigg)^{1-\de/2}\bigg). 
\eeqa
In view of (\ref{Cook-method-argument}), we get 
\beqa
\|\Psi_{t_2,\si_{t_2}}-\Psi_{t_1,\si_{t_1}}\|\leq c\sum_{i=1}^4\bigg(\fr{t_2^{1/3}}{t_1}\bigg)^{\eps_i}
\eeqa
for $0<\eps_i\leq 1$.
Let us now set $\Psi(t):=\Psi_{t,\si_t}$ and proceed as in the proof of Theorem~3.1 of \cite{Pi05}:
Suppose $t_1^n\leq t_2<t_1^{n+1}$. Then we can write
\beqa
\|\Psi(t_2)-\Psi(t_1)\|&\leq& \bigg(\sum_{k=1}^{n-1}\|\Psi(t_1^{k+1})-\Psi(t_1^k)\|\bigg)+\|\Psi(t_2)- \Psi(t_1^n)\|\non\\
&\leq&c\sum_{i=1}^4\sum_{k=1}^{n} \bigg( \fr{1}{t_1^{\eps_i(2k/3-1/3)}}  \bigg)\leq c\sum_{i=1}^4 t_1^{-\eps_i/3}\fr{ 1 }{1-(1/t_1)^{2\eps_i/3}}. \label{telescopic}
\eeqa
Since the last expression tends to zero as $t_1\to \infty$, we obtain  convergence of $t\mapsto \Psi(t)$. 

Finally, let us show that the limit $\Psi^+_{h_1,h_2}$ is independent of the choice of the parameter $\ga\in (4,\ga_0]$. Let $4<\ga'\leq \ga$
and let $\si_t=\ka/t^{\ga}$, $\si'_t=\ka/t^{\ga'}$ so that $\si_t\leq \si'_t$. We will show that
\beqa
\lim_{t\to \infty}\|\Psi_{t,\si_t}-\Psi_{t,\si'_t}\|=0. \label{difference-of-two-gammas}
\eeqa
Similarly as in the first part of the proof, it follows from formula~(\ref{tensor-product-structure})  
and from Proposition~\ref{preliminaries-on-spectrum}~\ref{eigenvectors-convergence}  that
\beqa
\lim_{t\to \infty}\lan \Psi_{t,\si_t}, \Psi_{t,\si_t}\ran=\lim_{t\to \infty}\lan \Psi_{t,\sip_t}, \Psi_{t,\si_t}\ran=
\lim_{t\to \infty}\lan \Psi_{t,\sip_t}, \Psi_{t,\sip_t}\ran
=\lan   \psi_{h_1},   \psi_{h_1}\ran \lan  \psi_{h_2}, \psi_{h_2}\ran.
\eeqa
This concludes the proof of (\ref{difference-of-two-gammas}). \qed\\
In Proposition~\ref{Cook-method-proposition} below we derive a formula for $\pa_t \Psi_{t,\si}$, appearing in assumption~(\ref{cook-method}).
For the purpose  of this derivation we introduce a suitable domain: First, we fix $l\in\nat_0$, $r_1,r_2>0$ and consider vectors of the form
\beqa
\Psi_l^{r_1,r_2}=\sum_{m=0}^{\infty}\fr{1}{\sqrt{m!}}\int d^{3l}p d^{3m}k\, F_{l,m}(p;k)\nn^*(p)^l\bb^*(k)^m\vac, \label{psi-l}
\eeqa
where $F_{l,m}\in L^2_{\sym}(\mcB_{r_1}^{\times l}\times\mcB_{r_2}^{\times m}, d^{3l}p d^{3m}k)$ i.e., 
$F_{l,m}$ are square-integrable functions, symmetric (independently) in their electron and photon variables and supported
in each electron (resp. photon) variable in a ball af radius $r_1$ (resp. $r_2$). 
Moreover, the norms of $F_{l,m}$ satisfy the bound
\beqa
\|F_{l,m}\|_2\leq \fr{c^m}{\sqrt{m!}} \label{factorial-bound}
\eeqa
for some $c\geq 0$, independent of $m$, which guarantees that the vector (\ref{psi-l}) is well defined. Now we set
\beqa
\mcD:=\mathrm{Span}\{\,  \Psi_l^{r_1,r_2} \,|\, l\in\nat_0, r_1,r_2>0\,\}, \label{invariant-domain}
\eeqa
where $\mathrm{Span}$ means finite linear combinations. This domain is dense and it contains $\mcC$. 
We show in Lemma~\ref{ren-creation-operator} that $H_{\si}, H_{\el}, H_{\pho}, H_{\I}^{\ain/\cin}$ and $\nr^*(h)$, $h\in \B(\PS)$, 
are well-defined  on $\mcD$ and leave this domain invariant. 
\bep\label{Cook-method-proposition} Let $h_1,h_2\in C_0^{2}(\PS)$ have disjoint supports and let $\Psi_{t,\si}$ 
be given by (\ref{two-par-approx}).  Then there holds
\beqa
\pa_t \Psi_{t,\si}&=&e^{itH}\bigg\{\h i[ [H_{\I}^{\ain}, \nr_{\si}^*(h_{1,t})  ], \nr_{\si}^*(h_{2,t})]\vac
+i \nr_{\si}^*(h_{1,t})\cH_{\I,\si}^{\cin} \nr_{\si}^*(h_{2,t})\vac\non\\
& &\pha{44444444444444444444444444}+i \nr_{\si}^*(h_{1,t}) \nr_{\si}^*(h^{\si}_{2,t})\vac\bigg\}+\{1\leftrightarrow 2\}. \label{time-derivative-terms}
\eeqa
The operators  $H_{\I}^{\ain}$ and $\cH_{\I}^{\cin}$ are defined on $\mcC$ by
\beqa
H_{\I}^{\ain}&:=&\int d^3p d^3k\,\vv(k)\nn^*(p+k)\bb(k)\nn(p), \\
\cH_{\I,\si}^{\cin}&:=&\int d^3p d^3k\,\cvv^{\si}(k)\nn^*(p-k)\bb^*(k)\nn(p), \label{check-hamiltonian}
\eeqa
where $\cvv^{\si}(k):=\g\fr{\mathbf{1}_{\mcB_\si}(k) |k|^{\alf} }{(2|k|)^{1/2}}$ and $h_{i}^{\si}(p)=(E_{p,\si}-E_p)h_{i}(p)$.
\eep
\proof We compute
\beqa
\pa_t \Psi_{t,\si}&=&e^{iHt}iH\nr_{\si}^*(h_{1,t})\nr_{\si}^*(h_{2,t})\vac+e^{iHt}\nr_{\si}^*(\pa_t h_{1,t})\nr_{\si}^*(h_{2,t})\vac
+e^{iHt}\nr_{\si}^*(h_{1,t})\nr_{\si}^*(\pa_t h_{2,t})\vac\non\\
&=&e^{iHt}iH\nr_{\si}^*(h_{1,t})\nr_{\si}^*(h_{2,t})\vac-e^{iHt}\nr_{\si}^*(h_{2,t})\nr_{\si}^*(i (Eh_1)_{t})\vac
-e^{iHt}\nr_{\si}^*(h_{1,t})\nr_{\si}^*(i(E h_2)_{t})\vac,    \non\\ \label{first-derivative-formula}
\eeqa
where $(Eh_i)(p):=E_ph_i(p)$, $i=1,2$.
The first term on the r.h.s. above is well defined by Lemma~\ref{ren-creation-operator}. 
The equality 
\beqa
(\pa_t\nr_{\si}^*( h_{1,t}))\nr_{\si}^*(h_{2,t})\vac=\nr_{\si}^*(\pa_t h_{1,t})\nr_{\si}^*(h_{2,t})\vac
\eeqa
can easily be justified with the help of Lemmas~\ref{norms-of-scattering-states} and \ref{summation}.
Now we note the following identity, which is meaningful due to Lemma~\ref{ren-creation-operator}:
\beqa
iH\nr_{\si}^*(h_{1,t})\nr_{\si}^*(h_{2,t})\vac&=&i[[H, \nr_{\si}^*(h_{1,t})],\nr_{\si}^*(h_{2,t})]\vac
+\nr_{\si}^*(h_{1,t})i H  \nr_{\si}^*(h_{2,t})\vac\non\\
& &+\nr_{\si}^*(h_{2,t})i H \nr_{\si}^*(h_{1,t})\vac. \label{H-double-commutator}
\eeqa
As for the first term on the r.h.s. of (\ref{H-double-commutator}), we note that $[H_{\free}, \nr_{\si}^*(h_{1,t})]$ and $[H_{\I}^{\cin}, \nr_{\si}^*(h_{1,t})]$
are sums of products of creation operators and therefore commute with $\nr_{\si}^*(h_{2,t})$. Thus we get
\beqa
i[[H, \nr_{\si}^*(h_{1,t})],\nr_{\si}^*(h_{2,t})]\vac=i[[H_{\I}^{\ain} , \nr_{\si}^*(h_{1,t})],\nr_{\si}^*(h_{2,t})]\vac.
\eeqa
As for the second  term on the r.h.s. of (\ref{H-double-commutator}), we obtain 
\beqa
\nr_{\si}^*(h_{1,t})i H  \nr_{\si}^*(h_{2,t})\vac&=&\nr_{\si}^*(h_{1,t})i (H-H_{\si}) \nr_{\si}^*(h_{2,t})\vac+\nr_{\si}^*(h_{1,t})\nr_{\si}^*(i(E_{\si}h_2)_{t})\vac\non\\
&=&\nr_{\si}^*(h_{1,t})i \cH_{\I,\si}^{\cin} \nr_{\si}^*(h_{2,t})\vac+\nr_{\si}^*(h_{1,t})\nr_{\si}^*(i((E_{\si}-E)h_2)_{t})\vac\non\\
& &+\nr_{\si}^*(ih_{1,t})\nr_{\si}^*( (Eh_2)_{t})\vac. \label{Lemma-enters}
\eeqa
Here in the first step we applied Lemma~\ref{fiber-ground-states-lemma} and in the last step we made use of the fact that
the operator
\beqa
\cH^{\ain}_{\I,\si}:=\int d^3p d^3k\,\cvv^{\si}(k)\nn^*(p+k)\bb(k)\nn(p)
\eeqa
annihilates $\nr_{\si}^*(h_{2,t})\vac$ due to the fact that $\cvv^{\si}$ is supported below the infrared cut-off. 
As the last term on the r.h.s. of (\ref{H-double-commutator}) can be treated analogously, this concludes the proof. \qed\\
\bel\label{fiber-ground-states-lemma} Let $h\in \B(\PS)$  and  let us consider the vector $\psi_{h,\si}:=\nr_{\si}^*(h)\vac$. 
Then
\beqa
\psi_{h,\si}= \Pi^*\int^{\oplus}dP\, h(P) \psi_{ P,\si}.
\eeqa
Consequently, $H_{\si} \psi_{h,\si}= \psi_{E_{\si}h,\si}$, where $(E_{\si}h)(p):=E_{p,\si}h(p)$.
\eel
\proof The $m$-particle components of $\psi_{h,\si}$, with the electron's variables  in the configuration space representation, 
have the form
\beqa
\psi_{h,\si}^{m}(x,k_1,\ldots,k_m)&=& \fr{1}{(2\pi)^{3/2}}\int d^3p\, e^{ipx} h(p+\unk) f^{m}_{p+\unk,\si }(k_1,\ldots, k_m )\non\\
&=& \fr{1}{(2\pi)^{3/2}} e^{-i\unk x }  \int d^3p\, e^{ipx} h(p) f^{m}_{p,\si }( k_1,\ldots, k_m).
\eeqa
Consequently, the $m$-particle  components of $\Pi( \psi_{h,\si})$ are
\beqa
\Pi(\psi_{h,\si})^{m}(P,k_1,\ldots k_m)&=& \fr{1}{(2\pi)^{3}} \int d^3x\, e^{-iP x }  \int d^3p\, e^{ipx} h(p) f^{m}_{p,\si }(k_1,\ldots, k_m)\non\\
&=& h(P) f^{m}_{P,\si }(k_1,\ldots, k_m),
\eeqa
which concludes the proof. \qed

\section{Vacuum expectation values of renormalized creation operators} \label{vacuum-expectation-values}
\setcounter{equation}{0}

In this section we estimate the norms of the terms appearing on the r.h.s. of (\ref{time-derivative-terms})
in order to verify assumption~(\ref{cook-method}) in Theorem~\ref{convergence-main-theorem}. We also
derive estimates on the norms of scattering states, required in this theorem. A crucial input is provided
by the non-stationary phase analysis in Section~\ref{non-stationary-phase}, which in turn relies on the spectral information from 
Theorem~\ref{main-theorem-spectral}, proven in \cite{DP12}.

In this section we will use the following definitions:
\beqa
G_{i,m}(q;k):=e^{-iE_qt}h_i(q)f^{m}_{q,\si}(k),\quad i\in\{1,2\}  \label{G-m-def-two}, 
\eeqa
where $h_1,h_2\in \B(\PS)$. (We stress that $G_{i,m}$ are $t$-dependent, although this is not reflected by the notation). 
Moreover, we set  
\beqa
B_{m}^*(G_{i,m}):=\int d^3q d^{3m}k\, G_{i,m}(q;k) \bb^*(k)^m \nn^*(q-\unk). \label{B-star-def}
\eeqa

Another convention, which we will use in this and the next  section, concerns contractions of creation and annihilation operators.
It is explained in full detail in Lemma~\ref{combinatorics}, so it is enough to illustrate it here by a simple example.
Let us set $\ti n=3, \ti m=2, n=2, m=3$ and consider photon variables $\ti r\in \real^{3\ti n}, \ti k\in \real^{3\ti m},  r\in \real^{3n}, k\in \real^{3 m}$.
Then the expectation value
 \beqa
& &\lan\vac, \bb(\ti r)^{\ti n} \bb(\ti k)^{\ti m} \bb^*(r)^{n} \bb^*(k)^{m}\vac\ran\non\\
& &=\lan\vac, \{\bb(\ti r_1) \bb(\ti r_2) \bb(\ti r_3)\} \, \{ \bb(\ti k_1)\bb(\ti k_2) \} \, \{  \bb^*(r_1)\bb^*(r_2) \}\, \{ \bb^*(k_1)\bb^*(k_2) \bb^*(k_3)\} \vac\ran
\eeqa
is a sum of $(m+n)!$ terms resulting from all the possible contraction patterns. Let us consider one of them: 
\beqa
\acontraction{ \lan\vac, \{  }{ \bb(\ti r_1)   }{ \bb(\ti r_2) \bb(\ti r_3)\} \, \{ \bb(\ti k_1)\bb(\ti k_2) \} \, \{   }{  \bb^*(r_1)   } 
\acontraction[2ex ]{ \lan\vac, \{\bb(\ti r_1)  } { \bb^*(r_2) } { \bb(\ti r_3)\} \, \{ \bb(\ti k_1)\bb(\ti k_2) \} \, \{  \bb^*(r_1)    }{ \bb^*(r_2)    } 
\acontraction[3ex ]{ \lan\vac, \{\bb(\ti r_1) \bb(\ti r_2)   }{   \bb(\ti r_3)  }{ \} \, \{ \bb(\ti k_1)\bb(\ti k_2) \} \, \{  \bb^*(r_1)\bb^*(r_2) \}\, \{  
  }{ \bb^*(k_1)   } 
\bcontraction{ \lan\vac, \{\bb(\ti r_1) \bb(\ti r_2) \bb(\ti r_3)\} \, \{    }{ \bb(\ti k_1)   }{ \bb(\ti k_2) \} \, \{  \bb^*(r_1)\bb^*(r_2) \}\, \{\bb^*(k_1)   }
{ \bb^*(k_2)   } 
\bcontraction[2ex ]{ \lan\vac, \{\bb(\ti r_1) \bb(\ti r_2) \bb(\ti r_3)\} \, \{ \bb(\ti k_1)   }{ \bb(\ti k_2)   }
{ \} \, \{  \bb^*(r_1)\bb^*(r_2) \}\, \{\bb^*(k_1)\bb^*(k_2)    }{  \bb^*(k_3)  } 
\lan\vac, \{\bb(\ti r_1) \bb(\ti r_2) \bb(\ti r_3)\} \, \{ \bb(\ti k_1) \bb(\ti k_2) \} \, \{  \bb^*(r_1)\bb^*(r_2) \}\, \{\bb^*(k_1)\bb^*(k_2) \bb^*(k_3)\} \vac\ran. \label{contraction-pattern}
\eeqa
Given this contraction pattern, we make a decomposition $\ti r=(\hat{\ti r}, \check{\ti r})$, where $\hat{\ti r}=(\ti r_1, \ti r_2)$ are the $\ti r$-variables which are contracted with some $r$-variables and $\check{\ti r}= \ti r_3 $ is the $\ti r$-variable contracted with a $k$-variable. In the 
case of the $\ti k$-variables we have $\ti k=\hat{\ti k}$, since both $\ti k_1$ and $\ti k_2$ are contracted with $k$-variables. In this
situation we say that $\check{\ti k }$ is empty.  Similarly, we have $r=\hat r$ with $\check r$ empty and $k=(\hat k, \check k)$ with
$\hat k=(k_2,k_3)$ and $\check k=\{k_1\}$.

\subsection{Double commutator}\label{double-commutator}

It turns out that the behaviour of the double commutator on the r.h.s. of (\ref{time-derivative-terms}) is governed by
the decay of the functions
\beqa
F_{n,m}^{G_1,G_2}(q; r \ba p; k):=(n+1)\int d^3r_{n+1}\, \vv(r_{n+1})G_{1,n+1}(q+r_{n+1} ; r,r_{n+1}) G_{2,m}(p-r_{n+1} ; k), \label{FGH}
\eeqa
where $q,p\in \real^3$, $r\in\real^{3n}$, $k\in\real^{3m}$.
For any such function we define an auxiliary operator
\beqa
B_{n,m}^*(F_{n,m}^{G_1,G_2}):=\int d^3q d^3p\int d^{3n}r  d^{3m}k\, F_{n,m}^{G_1,G_2}(q; r \ba p; k) \bb^*(r)^{n}\bb^*(k)^m\nn^*(p-\unk)\nn^*(q-\unr).
\label{double-creation}
\eeqa
Key properties of expectation values of such operators are given in Lemma~\ref{double-creation-lemma} below. Now we 
state and prove the estimate on the double commutator which gives rise to the first term on the r.h.s. of (\ref{cook-method}). 
\bep\label{double-commutator-proposition}  There holds the bound 
\beqa
\|[ [H_{\I}^{\ain}, \nr_{\si}^*(h_{1,t})  ], \nr_{\si}^*(h_{2,t})]\vac\|\leq  
\fr{c}{\si^{\de_{\g_0}}  } \bigg( \fr{\si^{\alf/(4\tiga)  } }{t}+\si t+\fr{1}{t^2\si^{1/(4\tiga)}} \bigg). 
\eeqa
\eep
\proof We compute
\beqa
& &\lan [ [H_{\I}^{\ain}, \nr_{\si}^*(h_{1,t})  ], \nr_{\si}^*(h_{2,t})]\vac,   [ [H_{\I}^{\ain}, \nr_{\si}^*(h_{1,t})  ], \nr_{\si}^*(h_{2,t})]\vac\ran\non\\
&=&\sum_{\substack{ m,n,\ti m,\ti n\in\nat_0 \\   m+n=\ti m+\ti n} }
\fr{1}{\sqrt{m!n!\ti m!\ti n!}} \lan  [[H_{\I}^{\ain},  B_{\ti n}^*(G_{1,\ti n})], B_{\ti m}^*(G_{2,\ti m})]  \vac,  [[H_{\I}^{\ain},  B_{n}^*(G_{1,n})], B_{m}^*(G_{2,m})]\vac\ran,\non\\
&=&\sum_{\substack{m,n,\ti m,\ti n \in \nat_0 \\  m+n=\ti m+\ti n} }
\fr{1}{\sqrt{m!n!\ti m!\ti n!}} \lan (B_{\ti n-1,\ti m}^*(F_{\ti n-1,\ti m}^{G_1,G_2})+B^*_{\ti m-1,\ti n}(F_{\ti m-1,\ti n}^{G_2,G_1}) ) \vac, \non\\
& &\pha{4444444444444444444444444444444} (B^*_{n-1,m}(F_{n-1,m}^{G_1,G_2})+B_{m-1,n}^*(F_{m-1,n}^{G_2,G_1}) )\vac\ran,
\label{scalar-product-formula-two}
\eeqa
where in the last step we made use of Lemma~\ref{double-comm-lemma} and the operators $B^*_{m,n}(\,\cdot\,)$ are defined in (\ref{double-creation}).
 Given families of functions $G_{i_1,n}, G_{i_2,n}, G_{j_1,n}, G_{j_2,n}$ of the form~(\ref{G-m-def-two}), with $i_1,i_2,j_1,j_2\in\{1,2\}$,
we define
\beqa
C(G_{i_1},G_{i_2};G_{j_1},G_{j_2})\pha{444444444444444444444444444444444444444444444444}\non\\
:=\sum_{\substack{m, n, \ti m, \ti n\in \nat_0 \\  m+n=\ti m+\ti n} }
\fr{1}{\sqrt{m!(n+1)!\ti m!(\ti n+1)!}}  \lan B_{\ti n,\ti m}^*(F_{\ti n,\ti m}^{G_{i_1},G_{i_2}})\vac, B_{n,m}^*(F_{n,m}^{G_{j_1},G_{j_2}})\vac\ran.
\label{C-definition}
\eeqa
Then we obtain from (\ref{scalar-product-formula-two}), taking into account  that $B_{-1,m}^*(F_{-1,m}^{G_1,G_2})\vac=B_{-1,n}^*(F_{-1,n}^{G_2,G_1})\vac=0$, 
\beqa
\|[ [H_{\I}^{\ain}, \nr_{\si}^*(h_{1,t})  ], \nr_{\si}^*(h_{2,t})]\vac\|^2=C(G_{1},G_2;G_1,G_2)&+&C(G_2,G_1;G_2,G_1)\non\\
&+&2\mathrm{Re}\big(C(G_1,G_2; G_2, G_1)\big).
\eeqa
 In view of definition~(\ref{G-m-def-two}),  $C(G_2,G_1;G_2,G_1)$ can
be obtained from $C(G_1,G_2;G_1,G_2)$ by a substitution $(h_1,h_2)\to (h_2,h_1)$. Thus it suffices to consider $C(G_1',G_2';G_1,G_2)$,
where
\beqa
G_{i,m}'(q;k)&:=&e^{-iE_qt}h_i'(q)f^{m}_{q,\si}(k),\quad i\in\{1,2\}
\eeqa
and $(h_1',h_2')\in\{(h_1,h_2), (h_2,h_1)\}$.  We recall from Lemma~\ref{double-creation-lemma} that
\beqa
& &\lan  B_{\ti n,\ti m}^*(F^{G_1',G_2'}_{\ti n,\ti m})\vac, B_{n,m}^*(F^{G_1,G_2}_{n,m})\vac\ran
=\sum_{\sig\in S_{m+n}}\int  d^3 q d^3p \int  d^{3n}r d^{3m}k\, F^{G_1,G_2}_{n,m}(q;r \ba p;k) \non\\
& &\times\bigg( \ov F_{\ti n,\ti m}^{G'_1,G'_2}(p-\uhk+\uhr; \hr, \chk \ba q+\uhk-\uhr; \hk, \chr )     
+ \ov F_{\ti n,\ti m}^{G'_1,G'_2}(q+\uchk-\uchr; \hr, \chk \ba p-\uchk+\uchr;\hk, \chr) \bigg). \label{double-commutator-exp}
\eeqa
The notation $\hk, \chk, \hr, \chr$ is explained  in Lemma~\ref{combinatorics} and at the beginning of this section. 
Now we obtain from Lemma~\ref{smooth-bounds} 
\beqa
|F^{G_1,G_2}_{n,m}(q; r \ba p; k)|\leq  \fr{1}{\si^{\de_{\g_0}}  }  \bigg( \fr{\si^{\alf/(4\tiga)  } }{t}+\si t+\fr{1}{t^2\si^{1/(4\tiga)}} \bigg)
 \fr{1}{\sqrt{m!n!} }D(p,q)g^m_{\si}(k) g^n_{\si}(r), \label{FGH-bound-one}
\eeqa
where $(p,q)\mapsto D(p,q)$ is a smooth, compactly supported function.  From this bound we get 
\beqa
& &|F^{G_1,G_2}_{n,m}(q;r \ba p; k)\ov F_{\ti n,\ti m}^{G'_1,G'_2}(p-\uhk+\uhr; \hr, \chk \ba q+\uhk-\uhr; \hk, \chr )|\non\\
& &\pha{4444}\leq  \fr{1}{\si^{\de_{\g_0}}  } \bigg( \fr{\si^{\alf/(4\tiga)  } }{t}+\si t+\fr{1}{t^2\si^{1/(4\tiga)}} \bigg)^2D'(p,q)
 \fr{1}{\sqrt{m!n!\ti m!\ti n!} }   g^m_{\si}(k)^2 g^n_{\si}(r)^2
\eeqa
and similarly
\beqa
& &|F^{G_1,G_2}_{n,m}(q; r \ba p ; k) \ov F_{\ti n,\ti m}^{G'_1,G'_2}(q+\uchk-\uchr; \hr, \chk \ba p-\uchk+\uchr; \hk, \chr)|\non\\
& &\pha{444}\leq  \fr{1}{\si^{\de_{\g_0}}  } \bigg( \fr{\si^{\alf/(4\tiga)  } }{t}+\si t+\fr{1}{t^2\si^{1/(4\tiga)}} \bigg)^2D'(p,q)
  \fr{1}{\sqrt{m!n!\ti m!\ti n!} }    g^m_{\si}(k)^2 g^n_{\si}(r)^2,
\eeqa
where $D'$ is again a smooth, compactly supported function. 
Making use of (\ref{C-definition}),(\ref{double-commutator-exp}) and the last two bounds we get
\beqa
& &|C(G_1',G_2';G_1,G_2)|\non\\
&\leq& \fr{1}{\si^{\de_{\g_0}}  } \bigg( \fr{\si^{\alf/(4\tiga)  } }{t}+\si t+\fr{1}{t^2\si^{1/(4\tiga)}} \bigg)^2
\sum_{\substack{m, n, \ti m, \ti n \in \nat_0 \\  m+n=\ti m+\ti n} }
\fr{(m+n)!}{\sqrt{m!(n+1)!\ti m!(\ti n+1)!}}\non\\
& &\pha{44444444444444444444444444444444444}\times \fr{1}{\sqrt{m!n!\ti m!\ti n!} }   \|g^m_{\si}\|_2^2 \|g^n_{\si}\|_2^2\non\\
& \leq& \fr{1}{\si^{\de_{\g_0}}  } \bigg( \fr{\si^{\alf/(4\tiga)  } }{t}+\si t+\fr{1}{t^2\si^{1/(4\tiga)}} \bigg)^2 \bigg(\fr{\kas}{\si}\bigg)^{4\g^2 c^2} 
 \leq \fr{c'}{\si^{\de'_{\g_0}}  } \bigg( \fr{\si^{\alf/(4\tiga)  } }{t}+\si t+\fr{1}{t^2\si^{1/(4\tiga)}} \bigg)^2,\,\,\,\,\,\,\,\,\,
\eeqa
where we made use of Lemma~\ref{summation} and definition~(\ref{def-g-ks}). This concludes the proof. \qed
\bel\label{double-comm-lemma} 
 For any  $n,m\in \nat_0$ there holds
\beqa
[[H_\I^{\ain},  B_{n}^*(G_{1,n})],B_{m}^*(G_{2,m})]\vac=B_{n-1,m}^*(F_{n-1,m}^{G_1,G_2})\vac+B_{m-1,n}^*(F_{m-1,n}^{G_2,G_1})\vac,\label{double-comm-form}
\eeqa
where $F_{n-1,m}^{G_1,G_2}$ is defined in (\ref{FGH}) and  we set $B_{-1,m}^*(F_{-1,m}^{G_1,G_2})\vac=B_{-1,n}^*(F_{-1,n}^{G_2,G_1})\vac=0$. 
\eel
\proof First we compute the inner commutator on  $\mcC$:
\beqa
[H_\I^{\ain},  B_{n}^*(G_{1,n})]=\int d^3q d^{3n}r d^3u d^3w\,  G_{1,n}(q;r)\vv(w)[ \bb(w)\nn^*(u+w)\nn(u), \bb^*(r)^n \nn^*(q-\unr)].\,\,
\eeqa
We note that
\beqa
[\bb(w)\nn^*(u+w)\nn(u), \bb^*(r)^n \nn^*(q-\unr)]
&=&\sum_{i=1}^n\de(w-r_i)\bb^*(r_{\ci})^{n-1}\nn^*(u+w)\nn(u)\nn^*(q-\unr)\non\\
& &+\bb(w)\bb^*(r)^n\de(u-q+\unr)\nn^*(u+w), \label{commutator-two-terms}
\eeqa 
where $\bb^*(r_{\ci})^{n-1}=\bb^*(r_1)\ldots \bb^*(r_{i-1})\bb^*(r_{i+1})\ldots \bb^*(r_n)$ for $n\geq 1$ and $\bb^*(r_{\ci})^{n-1}=0$ for $n=0$. Since $G_{1,n}$ is symmetric in the photon variables,
the contributions of the first and the second term on the r.h.s. of (\ref{commutator-two-terms}) are given by
\beqa
& &\,[H_\I^{\ain},  B_{n}^*(G_{1,n})]_1\non\\
& &:= n\int d^3q d^{3(n-1)}r d^3r_n d^3u \,  G_{1,n}(q;r,r_n)\vv(r_n)\bb^*(r)^{n-1}\nn^*(u+r_n)\nn(u)\nn^*(q-\unr-r_n),\,\,\,\,\,\,\,\,\,
\eeqa
and
\beqa
 [H_{\I}^{\ain},  B_{n}^*(G_{2,n})]_2:=\int d^3q d^{3n}r d^3w\,  G_{2,n}(q;r)\vv(w) \bb(w)\bb^*(r)^n\nn^*(q-\unr+w),
\eeqa
respectively.
Now let us compute the first contribution to the double commutator:
\beqa
& &[[H_\I^{\ain},  B_{n}^*(G_{1,n})]_1, B_{m}^*(G_{2,m})]\vac\non\\
& &\pha{444444444}=n\int d^3q d^{3(n-1)}r d^3r_n d^3u\int d^3p d^{3m}k\,  G_{1,n}(q;r,r_n)\vv(r_n) G_{2,m}(p;k)\non\\ 
& &\pha{44444444444}\times [\bb^*(r)^{n-1}\nn^*(u+r_n)\nn(u)\nn^*(q-\unr-r_n),\bb^*(k)^m \nn^*(p-\unk)]\vac\non\\
& &\pha{444444444}=n\int d^3q d^3p  \int d^{3(n-1)}r d^3r_n d^{3m}k\,  G_{1,n}(q;r,r_n)\vv(r_n) G_{2,m}(p;k)\non\\ 
& &\pha{44444444444}\times \bb^*(r)^{n-1}\bb^*(k)^m\nn^*(p-\unk+r_n)\nn^*(q-\unr-r_n)\vac.  \label{double-commutator-one}
\eeqa
By changing variables $p\to p-r_n$ and $q\to q+r_n$, we get 
\beqa
[[H_\I^{\ain},  B_{n}^*(G_{1,n})]_1, B_{m}^*(G_{2,m})]\vac=B_{n-1,m}^*(F_{n-1,m}^{G_1,G_2})\vac .
\eeqa
The second contribution to the double commutator has the form:
\beqa
& &[[ H_\I^{\ain} ,  B_{n}^*(G_{1,n})]_2, B_{m}^*(G_{2,m})]\vac\non\\
& &\pha{444444444}=\int d^3q d^{3n}r d^3w \int d^3p d^{3m}k\,  G_{1,n}(q;r)\vv(w)G_{2,m}(p;k)\non\\ 
& &\pha{44444444444}\times [\bb(w)\bb^*(r)^n\nn^*(q-\unr+w), \bb^*(k)^m \nn^*(p-\unk)]\vac\non\\
& &\pha{444444444}=m\int d^3q d^3p  \int d^{3n}r  d^{3(m-1)}k d^3k_m\,  G_{1,n}(q;r)\vv(k_m)G_{2,m}(p;k,k_m)\non\\ 
& &\pha{44444444444}\times \bb^*(r)^n\bb^*(k)^{m-1}\nn^*(q-\unr+k_m)  \nn^*(p-\unk-k_m)\vac. \label{double-commutator-two}
\eeqa
By changing variables $q\to q-k_m$ and $p\to p+k_m$ we obtain
\beqa
[[ H_\I^{\ain},  B_{n}^*(G_{1,n})]_2, B_{m}^*(G_{2,m})]\vac=B_{m-1,n}^*(F_{m-1,n}^{G_2,G_1})\vac,
\eeqa
which concludes the proof. \qed

\bel\label{summation} There hold the estimates
\beqa
& &\sum_{\substack{m,n,\ti m, \ti n\in\nat_0, \\ m+n=\ti m+\ti n} }  \fr{(\ti m+\ti n)!}{m!n! \ti m!\ti n! } \|g^m_{\si}\|_2^2 \,\|g^n_{\si}\|_2^2
\leq \bigg(\fr{\kas}{\si}\bigg)^{4\la^2 c^2},\\
& &\sum_{\substack{m, n, \ti m, \ti n\in\nat_0 \\ m+n=\ti m+\ti n }}  \fr{(\ti m+\ti n)!}{m!n! \ti m!\ti n! } \|g^{m-1}_{\si}\|_2^2 \,\|g^n_{\si}\|_2^2
\leq 2 \bigg(\fr{\kas}{\si}\bigg)^{4\la^2 c^2},  \label{shifted-index}\\
& &\sum_{\substack{m, n, \ti m, \ti n\in\nat_0 \\ m+n=\ti m+\ti n }}  \fr{(\ti m+\ti n)!}{m!n! \ti m!\ti n! } \|g^{m-1}_{\si}\|_2^2 \,\|g^{n-1}_{\si}\|_2^2
\leq 4 \bigg(\fr{\kas}{\si}\bigg)^{4\la^2 c^2},  \label{shifted-second-index}
\eeqa
where $g^m_{\si}, \kas$ are defined in (\ref{def-g-ks}) and we set by convention $g^{-1}_{\si}=0$.
\eel
\proof By definition of the functions $g^m_{\si}$
\beqa
\|g^m_{\si}\|_2^2\leq (\g  c)^{2m}(\log(\kas/\si))^m.
\eeqa
Thus we get
\beqa
& &\sum_{\substack{m, n, \ti m,\ti n\in\nat_0 \\  m+n=\ti m+\ti n} }  \fr{(\ti m+\ti n)!}{m!n! \ti m!\ti n! }  \|g^m_{\si}\|_2^2 \|g^n_{\si}\|_2^2
 \leq \sum_{m,n\in\nat_0}\bigg(\sum_{ \substack{\ti m,\ti n\in\nat_0 \\ \ti m+\ti n=m+n}}  \fr{(\ti m+\ti n)!}{ \ti m!\ti n!}\bigg) \fr{ (\g  c)^{2(m+n)} }{m!n!}   (\log(\kas/\si))^{m+n}\non\\
& &\pha{4444444444444444444444444}\leq \sum_{m,n\in\nat_0} \fr{ (\sqrt{2}\la  c)^{2(m+n)} }{m!n!}   (\log(\kas/\si))^{m+n}
=\bigg(\fr{\kas}{\si}\bigg)^{4\g^2 c^2}.\,\,\,\,\,\,\,\quad
\eeqa
Let us now prove (\ref{shifted-index}):
\beqa
& &\sum_{\substack{m, n, \ti m,\ti n\in\nat_0 \\  m+n=\ti m+\ti n} }  \fr{(\ti m+\ti n)!}{m!n! \ti m!\ti n! }  \|g^{m-1}_{\si}\|_2^2 \|g^n_{\si}\|_2^2
\leq \sum_{ \substack{m, n,\ti m,  \ti n\in\nat_0 \\  m+n+1=\ti m+\ti n } }  \fr{(\ti m+\ti n)!}{m!n! \ti m!\ti n! }  \|g^{m}_{\si}\|_2^2 \|g^n_{\si}\|_2^2 \non\\
& &\pha{4444444444444444444444}\leq \sum_{m,n\in\nat_0}\bigg(\sum_{ \substack{ \ti m,\ti n\in\nat_0 \\ \ti m+\ti n=m+n+1} }  \fr{(\ti m+\ti n)!}{ \ti m!\ti n!}\bigg) \fr{ (\g  c)^{2(m+n)} }{m!n!}   (\log(\kas/\si))^{m+n}\non\\
& &\pha{4444444444444444444444}\leq2 \sum_{m,n\in\nat_0} \fr{ (\sqrt{2}\la  c)^{2(m+n)} }{m!n!}   (\log(\kas/\si))^{m+n}=2\bigg(\fr{\kas}{\si}\bigg)^{4\g^2 c^2}.\quad\quad
\eeqa
The bound (\ref{shifted-second-index}) is proven analogously. This concludes the proof. \qed

\subsection{Clustering of scalar products}\label{clustering-subsection}

In this subsection we consider clustering properties of scalar products of scattering states approximants.
We will study  the expression 
\beqa
\lan\Psi_{t,\sip}',\Psi_{t,\si}\ran= \lan \Om, \nr_{\sip}(h'_{2,t}) \nr_{\sip}(h'_{1,t})  \nr_{\si}^*(h_{1,t})\nr_{\si}^*(h_{2,t})\vac\ran.
\label{scalar-product-schematic}
\eeqa
Recalling from (\ref{renormalized}) that the renormalized creation operators have the form
\beqa
\nr_{\si}^*(h)=\sum_{m=0}^{\infty}\fr{1}{\sqrt{m!}}\int d^3p\,d^{3m}k\,h(p) f^{m}_{p,\si}(k)\bb^*(k)^m \, \nn^*(p- \unk),
\eeqa
we obtain that (\ref{scalar-product-schematic}) is a sum of integrals over vacuum expectation values of the electron and photon  
creation and annihilation operators. Contractions of the electron operators give rise to two terms
\beqa
\acontraction{\lan\Psi_{t,\sip}',\Psi_{t,\si}\ran^{(1)}:=\lan \Om, }{\nr_{\sip}(h'_{2,t}) }{ \nr_{\sip}(h'_{1,t})  \nr_{\si}^*(h_{1,t}) }{ \nr_{\si}^*(h_{2,t}) }
\acontraction[2ex]{\lan\Psi_{t,\sip}',\Psi_{t,\si}\ran^{(1)}:=\lan \Om, \nr_{\sip}(h'_{2,t})  }{ \nr_{\sip}(h'_{1,t})   }{   }{ \nr_{\si}^*(h_{1,t}) }
\lan\Psi_{t,\sip}',\Psi_{t,\si}\ran^{(1)}:=\lan \Om, \nr_{\sip}(h'_{2,t}) \nr_{\sip}(h'_{1,t})  \nr_{\si}^*(h_{1,t})\nr_{\si}^*(h_{2,t})\vac\ran,\label{direct}\\
\acontraction{ \lan\Psi_{t,\sip}',\Psi_{t,\si}\ran^{(2)}:=\lan \Om,   }{ \nr_{\sip}(h'_{2,t})  }{ \nr_{\sip}(h'_{1,t})  }{ \nr_{\si}^*(h_{1,t}) }
\acontraction[2ex]{ \lan\Psi_{t,\sip}',\Psi_{t,\si}\ran^{(2)}:=\lan \Om, \nr_{\sip}(h'_{2,t})  }{ \nr_{\sip}(h'_{1,t}) }{ \nr_{\si}^*(h_{1,t}) }{ \nr_{\si}^*(h_{2,t}) }
\lan\Psi_{t,\sip}',\Psi_{t,\si}\ran^{(2)}:=\lan \Om, \nr_{\sip}(h'_{2,t}) \nr_{\sip}(h'_{1,t})  \nr_{\si}^*(h_{1,t})\nr_{\si}^*(h_{2,t})\vac\ran, \label{exchange}
\eeqa
which we call `direct' and `exchange', respectively. We emphasize that the contractions in (\ref{direct}) and (\ref{exchange}) above do not involve 
the photon creation and annihilation operators. The contractions of photon variables are the subject of the remaining part of this discussion. 

As for the direct term, we distinguish two types  of photon contraction
patterns. The first type has the form
\beqa
\acontraction{\lan\vac, }{ \bb(\ti k)^{\ti m}   }{ \bb(\ti r)^{\ti n}  \bb^*(r)^{n}  }{ \bb^*(k)^{m} }
\acontraction[2ex]{ \lan\vac,\bb(\ti k)^{\ti m} }{ \bb(\ti r)^{\ti n} }{    }{ \bb^*(r)^{n}}
\lan\vac, \bb(\ti k)^{\ti m} \bb(\ti r)^{\ti n}  \bb^*(r)^{n} \bb^*(k)^{m}\vac\ran, \label{direct-photon-contraction-pattern}
\eeqa
where the monomials of creation and annihilation operators above come from the respective renormalized creation operators 
in (\ref{direct}).   By (\ref{direct-photon-contraction-pattern}) we mean that  all the $\ti k$-variables are contracted with $k$-variables
and all the $\ti r$-variables are contracted with $r$-variables, and that $\ti m=m$, $\ti n=n$. (That is, $\check{k}$ and $\check{r}$
are empty in the terminology introduced below equation~(\ref{contraction-pattern})). 
Clearly, for fixed $m, n$ there are $m!n!$ such contraction
patterns and, as we will show in the proof  of Proposition~\ref{scalar-product-proposition}, after summation over $m,n$ they give rise to the first term on the r.h.s. of (\ref{scalar-product-expression}),
which is $\lan  \psi_{h_1',\sip},  \psi_{h_1,\si}\ran \lan  \psi_{h_2',\sip}, \psi_{h_2,\si}\ran$.  Contraction patterns of the second type
are those for which  $\check{ k}$ or $\check{ r}$ are non-empty. It will be shown with the help of the non-stationary phase analysis from 
Lemma~\ref{second-contribution-decay} that the resulting terms contribute to the rest term $R(t,\si,\si')$ on the r.h.s. of (\ref{scalar-product-expression}),  which  eventually tends to zero in the proof of Theorem~\ref{convergence-main-theorem}.

As for the exchange term (\ref{exchange}), we consider again two types of photon contraction patterns. Contractions of the first
type have the form
\beqa
\acontraction{\lan\vac,  }{ \bb(\ti k)^{\ti m}  }{ \bb(\ti r)^{\ti n}   }{ \bb^*(r)^{n}  }
\acontraction[2ex]{ \lan\vac, \bb(\ti k)^{\ti m}   }{ \bb(\ti r)^{\ti n} }{  \bb^*(r)^{n}  }{ \bb^*(k)^{m}  }
\lan\vac, \bb(\ti k)^{\ti m} \bb(\ti r)^{\ti n}  \bb^*(r)^{n} \bb^*(k)^{m}\vac\ran, \label{exchange-photon-contraction-pattern}
\eeqa 
i.e., $\hat{r}$ and $\hat{k}$ are empty. These contraction patterns give rise to the second term on the r.h.s.   (\ref{scalar-product-expression}),
which is $\lan  \psi_{h_1',\sip},  \psi_{h_2,\si}\ran \lan  \psi_{h_2',\sip}, \psi_{h_1,\si}\ran$. Contraction patterns of the second type are
those for  which $\hat r$ or $\hat k$ are non-empty. The non-stationary phase analysis from Lemma~\ref{First-contribution-decay} gives
that these terms contribute to $R(t,\si,\si')$.

After this overview, we are ready to state and prove the main result of this subsection which yields estimate~(\ref{tensor-product-rest-term}).
\bep\label{scalar-product-proposition} Let $\ka\geq \sip\geq \si>0$ and let $\Psi_{t,\si}$, $\Psi'_{t,\si}$ be two scattering states approximants 
 given by (\ref{two-par-approx}) with momentum wave functions $h_1,h_2$ and $h_1',h_2'$, respectively. (We recall that 
$\supp\, h_1\cap \supp\, h_2=\emptyset$ and $\supp\, h_1'\cap \supp\, h_2'=\emptyset$. However, $\supp\, h_i\cap \supp\, h_j'$, 
$i, j\in\{1,2\}$ may be non-empty).
Then
\beqa
\lan\Psi_{t,\sip}',\Psi_{t,\si}\ran&=&\lan  \psi_{h_1',\sip},  \psi_{h_1,\si}\ran \lan  \psi_{h_2',\sip}, \psi_{h_2,\si}\ran\non\\
& &+ \lan  \psi_{h_1',\sip},  \psi_{h_2,\si}\ran \lan  \psi_{h_2',\sip}, \psi_{h_1,\si}\ran +R(t,\si,\si'), \label{scalar-product-expression}
\eeqa
where the rest term satisfies
\beqa
|R(t,\si,\si')|\leq  C(h,h')\fr{1}{\si^{\de_{\g_0} }} \bigg( \fr{1}{t}\fr{1}{\si^{1/(8\tiga)} }+ (\si')^{2\alf}  \bigg).   
\eeqa
Here $C(h,h'):=c\|h_1\|_1\|h_2\|_1 \sum_{\be_1,\be_2; 0\leq |\be_1|+|\be_2|\leq 1} \|\pa^{\be_1}h_1'\|_{\infty} \|\pa^{\be_2}h_2'\|_{\infty}$
and the sum extends over multiindices $\be_1$, $\be_2$.
\eep
\proof Let us set
\beqa
G_{i,m}(q;k)&:=&e^{-iE_qt}h_i(q)f^{m}_{q,\si}(k), \label{G-def-repeated}\\
G'_{i,m}(q;k)&:=&e^{-iE_qt}h_i'(q)f^{m}_{q,\si'}(k), \label{G-prime-def}
\eeqa
for  $i\in\{1,2\}$.
Now we can write
\beqa
\lan\Psi_{t,\sip}',\Psi_{t,\si}\ran=
\sum_{\substack{m,n,\ti m,\ti n\in\nat_0 \\ \ti m+\ti n=m+n } }\fr{1}{\sqrt{m!n!\ti m!\ti n!}} 
\lan\vac, B_{\ti n}(G_{1,\ti n}' )B_{\ti m}(G_{2,\ti m}' )B_{n}^*(G_{1,n}) B_{m}^*(G_{2,m})\vac\ran.\,\,
\label{scalar-product-formula}
\eeqa
Making use of Lemma~\ref{norms-of-scattering-states}, we obtain
\beqa
\lan\vac, B_{\ti n}(G_{1,\ti n}' )B_{\ti m}(G_{2,\ti m}' )B_{n}^*(G_{1,n}) B_{m}^*(G_{2,m})\vac\ran \pha{4444444444444444444444444444444}  \non\\
=\sum_{\sig\in S_{m+n}}\int  d^3 q d^3p \int  d^{3n}r d^{3m}k\,  
G_{1,n}(q;  r) G_{2,m}(p;  k)\pha{444444444444444444444444}\non\\
 \times\bigg(\ov G_{1,\ti n}'( q+\un{\check{k}}-\un{\check{r}}; \hat r, \check k)\ov  G_{2,\ti m}'(p-\un{\check{k}}+\un{\check{r}}; \hat k, \check r)
+\ov G_{1,\ti n}'(p-\un{\hat k}+\un{\hat r}; \hat r, \check k) \ov G_{2,\ti m}'(q+\un{\hat{k}}-\un{\hat{r}}; \hat k, \check r)\bigg),
\label{B-expectation-value}
\eeqa
where the notation in (\ref{B-expectation-value}) is explained in Lemma~\ref{combinatorics}.

Let us denote the summands involving the first term in the bracket on the r.h.s. of (\ref{B-expectation-value}) by $I^{(1)}_{m,n,\ti m,\ti n}$.
Let $\check I^{(1)}_{m,n,\ti m,\ti n}$ be such summands coming from permutations for which $\check k$
or $\check r$ are non-empty (cf. the discussion below formula~(\ref{contraction-pattern})). We note that there are $(m+n)!-m!n!$  such permutations. 
Let $\lan\Psi'_{t,\sip},\Psi_{t,\si}\ran^{(\check 1)}$
be the  contribution to (\ref{scalar-product-formula}) involving all the summands $\check I^{(1)}_{m,n,\ti m,\ti n}$. Making use of Lemmas~\ref{second-contribution-decay} and  \ref{summation}, we get
\beqa
|\lan\Psi_{t,\sip}',\Psi_{t,\si}\ran^{(\check 1)}|&\leq& 
C(h,h') \fr{1}{\si^{\de_{\g_0} }} \bigg( \fr{1}{t}\fr{1}{\si^{1/(8\tiga)} }+ (\si')^{2\alf}  \bigg) \bigg(\fr{\kas}{\si}\bigg)^{4\la^2 c^2}\non\\
&\leq& C(h,h')\fr{1}{\si^{\de'_{\g_0} }} \bigg( \fr{1}{t}\fr{1}{\si^{1/(8\tiga)} }+ (\si')^{2\alf}  \bigg),
\eeqa
Clearly, this expression contributes to $R(t,\si,\si')$.

Now let $\hat I^{(1)}_{m,n,\ti m,\ti n}$ denote the expressions $I^{(1)}_{m,n,\ti m,\ti n}$ coming from permutations for which $\check k$ and $\check r$
are empty. (We note that there are $m!n!$ such permutations and that in this case $m=\ti m$, $n=\ti n$). We obtain from 
Lemma~\ref{norms-of-single-particle-states} that
\beqa
\hat I^{(1)}_{m,n,\ti m,\ti n}&=&\int  d^3 q  d^{3n}r \,  G_{1,n}(q;  r) \ov G_{1,\ti n}'(q;  r)  \int d^3p  d^{3m}k\,   
G_{2,m}(p;  k ) \ov  G_{2,\ti m}'(p;  k)\non\\
&=&\fr{1}{n!}\fr{1}{m!}\lan \vac, B_{\ti n}(G_{1,\ti n}') B_n^*(G_{1,n})\vac\ran  \lan \vac,B_{\ti m}(G_{2,\ti m}') B_m^*(G_{2,m})\vac\ran. 
\eeqa
Let $\lan\Psi_{t,\sip}',\Psi_{t,\si}\ran^{(\hat 1)}$ be the contribution to $\lan\Psi'_{t,\sip},\Psi_{t,\si}\ran$ involving all such  
$\hat I^{(1)}_{m,n,\ti m,\ti n}$. Since the sum over permutations in (\ref{B-expectation-value})  gives the compensating factor $m!n!$, we obtain
\beqa
\lan\Psi'_{t,\sip},\Psi_{t,\si}\ran^{(\hat 1)}&=&\sum_{m,n,\ti m,\ti n  }\fr{1}{\sqrt{m!n!\ti m!\ti n!}} 
\lan \vac, B_{\ti n}(G_{1,\ti n}') B_n^*(G_{1,n})\vac\ran  \lan \vac,B_{\ti m}(G_{2,\ti m}') B_m^*(G_{2,m})\vac\ran\non\\
&=&\lan   \psi_{h_1',\sip},   \psi_{h_1,\si}\ran 
\lan  \psi_{h_2',\sip}, \psi_{h_2,\si}\ran,
\eeqa
where in the last step we compared  definition~(\ref{B-star-def}) of  $B_n^*(G_{i,n})$ with definition~(\ref{renormalized})
of the renormalized creation operator and recalled that $\psi_{h_i,\si}=\nr^*_{\si}(h_i)\Om$.

The analysis of the  summands involving the second term in bracket on the r.h.s. of (\ref{B-expectation-value}) is analogous, so 
we can be brief: Let us denote such summands by $I^{(2)}_{m,n,\ti m,\ti n}$ and let
 $\hat I^{(2)}_{m,n,\ti m,\ti n}$ be the summands coming from permutations for which $\hk$
or $\hr$ are non-empty. We denote by $\lan\Psi'_{t,\sip},\Psi_{t,\si}\ran^{(\hat 2)}$
the  contribution to (\ref{scalar-product-formula}) involving all $\hat I^{(2)}_{m,n,\ti m,\ti n}$. By
Lemmas~\ref{First-contribution-decay} and  \ref{summation}, we get
\beqa
|\lan\Psi'_{t,\sip},\Psi_{t,\si}\ran^{(\hat 2)}|\leq  C(h,h')\fr{1}{\si^{\de'_{\g_0} }} \bigg( \fr{1}{t}\fr{1}{\si^{1/(8\tiga)} }+ (\si')^{2\alf}  \bigg). 
\eeqa
This part contributes to the rest term.

Now let $\check I^{(2)}_{m,n,\ti m,\ti n}$ be the expressions $I^{(2)}_{m,n,\ti m,\ti n}$ coming from permutations for which $\hk$ and $\hr$
are empty.  Denoting the corresponding contribution to  (\ref{scalar-product-formula}) by $\lan\Psi'_{t,\sip},\Psi_{t,\si}\ran^{(\check 2)}$, 
we get from Lemma~\ref{norms-of-single-particle-states} that
\beqa
\lan\Psi'_{t,\sip},\Psi_{t,\si}\ran^{(\check 2)}=\lan   \psi_{h_1',\sip},   \psi_{h_2,\si}\ran 
\lan  \psi_{h_2',\sip}, \psi_{h_1,\si}\ran.
\eeqa
This concludes the proof.  \qed\\
The above proposition has a simple corollary which gives the last term on the r.h.s. of estimate~(\ref{cook-method}).
\bec\label{h-sigma-part} Let $h_1,h_2\in C_0^2(\PS)$ have disjoint supports and let $h^{\si}_2(q)=(E_{q,\si}-E_q)h_2(q)$. Then 
\beqa
\|\nr_{\si}^*(h_{1,t}) \nr_{\si}^*(h^{\si}_{2,t})\vac\|\leq c\si^{1/2-\de_{\la_0}}\bigg( 1+ \fr{1}{t}\fr{1}{\si^{1/(8\tiga)} }  \bigg).
\eeqa
\eec
\proof Making use of Proposition~\ref{scalar-product-proposition}, we get
\beqa
\|\nr_{\si}^*(h_{1,t}) \nr_{\si}^*(h^{\si}_{2,t})\vac\|^2=\lan  \psi_{h_1,\si},  \psi_{h_1,\si}\ran \lan  \psi_{h_2^{\si},\si}, \psi_{h_2^{\si},\si}\ran
+R(t,\si,\si),
\eeqa
where $R(t,\si,\si)$ satisfies 
\beqa
|R(t,\si,\si)|&\leq& C(h,h')\fr{1}{\si^{\de_{\g_0} }} \bigg( \fr{1}{t}\fr{1}{\si^{1/(8\tiga)} }+ (\si)^{2\alf}  \bigg), \\  
C(h,h)&=&c\|h_1\|_1\|h_2^{\si}\|_1 \sum_{\be_1,\be_2; 0\leq |\be_1|+|\be_2|\leq 1} \|\pa^{\be_1}h_1\|_{\infty} \|\pa^{\be_2}h_2^{\si}\|_{\infty}.
\eeqa
Making use of Proposition~\ref{preliminaries-on-spectrum}, we obtain that $C(h,h)\leq c'\si$ and therefore
\beqa
|R(t,\si,\si)|\leq c\si^{1-\de_{\la_0}}\bigg(1+\fr{1}{t}\fr{1}{\si^{1/(8\tiga)} }  \bigg).
\eeqa
Now making use of Lemma~\ref{fiber-ground-states-lemma},  we get $\|\psi_{h_1,\si}\|^2=\|h_1\|_2^2$ and 
$\|\psi_{h_2^{\si},\si}\|^2=\|h_2^{\si}\|_2^2$.  Exploiting  Proposition~\ref{preliminaries-on-spectrum} again, we get 
$\|\psi_{h_2^{\si},\si}\|^2\leq c\si^2$. Consequently, due to the constraint on the supports of $h_1$ and $h_2$, we obtain
\beqa
\|\nr_{\si}^*(h_{1,t}) \nr_{\si}^*(h^{\si}_{2,t})\vac\|^2\leq c'\si^{1-\de_{\la_0}}\bigg( 1+ \fr{1}{t}\fr{1}{\si^{1/(8\tiga)} }  \bigg).
\eeqa
This concludes the proof. \qed
\subsection{Contribution involving $\cH_{\I, \si}^{\cin}$}\label{single-commutator-section}
This subsection is devoted to the term involving $\cH_{\I, \si}^{\cin}$ on the r.h.s. of (\ref{time-derivative-terms}). The following
elementary proposition, which relies on Lemma~\ref{norms-of-scattering-states-one}, gives the second term on the r.h.s. of (\ref{cook-method}).
\bep\label{check-contribution} Let $\cH_{\I,\si}^{\cin}$ be  defined as in  (\ref{check-hamiltonian}). Then there holds the bound 
\beqa
\|\nr_{\si}^*(h_{1,t}) \cH_{\I,\si}^{\cin} \nr_{\si}^*(h_{2,t}) \vac\|\leq c\si^{1-\de_{\g_0}}.
\eeqa
\eep
\proof We rewrite the expression from the statement of the proposition as follows:
\beqa
   \lan \nr_{\si}^*(h_{1,t})  \cH_{\I,\si}^{\cin} \nr_{\si}^*(h_{2,t})   \vac, \nr_{\si}^*(h_{1,t}) \cH_{\I,\si}^{\cin} \nr_{\si}^*(h_{2,t})  \vac\ran \pha{444444444444444444444444444}  \non\\
=\sum_{\substack{m,n,\ti m,\ti n\in\nat_0 \\  m+n=\ti m+\ti n} }
\fr{1}{\sqrt{m!n!\ti m!\ti n!}} \lan\vac, B_{\ti n}(G_{2,\ti n}) (\cH_{\I,\si}^{\cin} )^*  B_{\ti m}(G_{1,\ti m} )
B_{n}^*(G_{1,n}) \cH_{\I,\si}^{\cin}  B_{m}^*(G_{2,m})\vac\ran.
\label{scalar-product-formula-one}
\eeqa
Now Lemma~\ref{norms-of-scattering-states-one} gives
\beqa
& &\lan\vac, B_{\ti n}(G_{2,\ti n} ) ( \cH_{\I,\si}^{\cin} )^*  B_{\ti m}(G_{1,\ti m} )B_{n}^*(G_{1,n}) \cH_{\I,\si}^{\cin}  B_{m}^*(G_{2,m})\vac\ran\non\\
&=&\sum_{\sig\in S_{m+n}}\int   d^3 q d^3p \int    d^{3n}r d^{3m}k\, G_{1,n}(q;r) G_{2,m}(p; k)\non\\
& &\times\bigg( \int d^3\ti p\, \cvv^{\si}(\ti p)^2  \ov G_{2,\ti n}(p-\ti p-\uhk+\uhr;\hr,\chk) \ov G_{1,\ti m}(\ti p+q+\uhk-\uhr; \hk,\chr)\non\\ 
& &\pha{444444444444444444444}+\|\cvv^{\si}\|_2^2  \ov G_{2,\ti n}(q+\un{\chk}-\un{\chr};\hr,\chk) \ov G_{1,\ti m}(p-\un{\chk}+\un{\chr}; \hk,\chr) \bigg).  
\label{H-check-auxiliary-one}
\eeqa
Making use of  Theorem~\ref{main-theorem-spectral}, and of definition~(\ref{G-m-def-two}), we obtain the bounds
\beqa
& &|G_{1,n}(q;r) G_{2,m}(p; k) \ov G_{2,\ti n}(p-\ti p-\uhk+\uhr;\hr,\chk) \ov G_{1,\ti m}(\ti p+q+\uhk-\uhr; \hk,\chr)|\non\\
& &\pha{44444444444444444444444444444}\leq  \fr{1}{\sqrt{m!n!\ti m!\ti n!}} D(p,q) g_{\si}^n(r)^2g_{\si}^m(k)^2,\\
& &|G_{1,n}(q;r) G_{2,m}(p; k) \ov G_{2,\ti n}(q+\un{\chk}-\un{\chr};\hr,\chk) \ov G_{1,\ti m}(p-\un{\chk}+\un{\chr}; \hk,\chr)|\non\\
& &\pha{44444444444444444444444444444}\leq \fr{1}{\sqrt{m!n!\ti m!\ti n!}}    D(p,q) g_{\si}^n(r)^2g_{\si}^m(k)^2,
\eeqa
 where $(p,q)\mapsto D(p,q)$ is some smooth compactly supported function independent of $\si$, $t$. 
Consequently, the r.h.s. of (\ref{H-check-auxiliary-one}) can be estimated by
\beqa
c\|\cvv^{\si}\|_2^2(m+n)! \fr{1}{\sqrt{m!n!\ti m!\ti n!}}   \|g_{\si}^n\|^2_2\|g_{\si}^m\|^2_2.
\eeqa
Substituting this bound to (\ref{scalar-product-formula-one}) and making use of Lemma~\ref{summation}, we get
\beqa
\|\nr_{\si}^*(h_{1,t}) \cH_{\I,\si}^{\cin} \nr_{\si}^*(h_{2,t}) \vac\|^2\leq c\|\cvv^{\si}\|_2^2\bigg(\fr{\kas}{\si}\bigg)^{4\la^2 c^2}.
\eeqa
Exploiting the fact that $\|\cvv^{\si}\|_2\leq c\si$, we conclude the proof. \qed


\section{Non-stationary phase arguments}\label{non-stationary-phase}
\setcounter{equation}{0}
In this section we derive  non-stationary phase estimates  which entered into our analysis in Subsections~\ref{double-commutator} 
and \ref{clustering-subsection}. The spectral information from Proposition~\ref{preliminaries-on-spectrum} 
and Theorem~\ref{main-theorem-spectral} is crucial for  this part of our investigation. 
\bel\label{smooth-bounds} Let $G_{i, m}$, $i\in \{1,2\}$, be as specified in  (\ref{G-m-def-two}) and let $F^{G_1,G_2}_{n,m}$ be defined as in (\ref{FGH}) i.e., it has the form
\beqa
F^{G_1,G_2}_{n,m}(q;r \ba p;k)
=(n+1)\int d^3\ti r\, \vv(\ti r) e^{-i(E_{ q+\ti r }+E_{p-\ti r} )t}h_1(p-\ti r )h_2(q+\ti r )
f^{n+1}_{q+\ti r ,\si}(r,\ti r ) f^{m}_{p-\ti r,\si}(k),\, \label{factor-n+1}
\eeqa
where $h_1,h_2\in C_0^{2}(\PS)$ have disjoint supports.
There holds the bound
\beqa
 |F^{G_1,G_2}_{n,m}(q; r \ba p; k)|\leq  \fr{1}{\si^{\de_{\g_0}}  }  \bigg( \fr{\si^{\alf/(4\tiga)  } }{t}+\si t+\fr{1}{t^2\si^{1/(4\tiga)}} \bigg)
\fr{1}{\sqrt{m!n!}}  D(p,q)g^m_{\si}(k) g^n_{\si}(r), \label{FGH-bound}
\eeqa
where  $D\in C_0^{\infty}(\real^3\times \real^3)$.  
\eel
\proof Let us introduce the slow cut-off $\si_\s:=\ka(\si/\ka)^{1/(8\tiga)}$, which clearly satisfies $\si\leq\si_\s\leq \ka$.
Let $\chi\in C^{\infty}(\real^3)$, $0\leq \chi\leq 1$,  be supported in $\mcB_1$ (the unit ball) and be equal to one on $\mcB_{1-\eps}$ for
some small $0<\eps<1$. We set $\chi_1(\ti k)=\chi(\ti k/\si_\s)$, $\chi_2(\ti k)=1-\chi_1(\ti k)$ and define
 \beqa
& &F^{G_1,G_2}_{j,n,m}(q; r \ba p; k)
:=\int d^3\ti r\, \vv(\ti r) \chi_j(\ti r) e^{-i(E_{ q+\ti r }+E_{p-\ti r})t }h_1(p-\ti r )h_2(q+\ti r )\non\\
& &\pha{4444444444444444444444444444444444444444} \times f^{n+1}_{q+\ti r ,\si}(r,\ti r ) f^{m}_{p-\ti r,\si}(k).   \label{decomposition-of-FGH}
\eeqa
We set $\nF(q,p,\ti r):=E_{ q+\ti r }+E_{p-\ti r }$ and note that by disjointness of the velocity
supports of $h_1$, $h_2$, the condition $h_1(p-\ti r )h_2(q+\ti r )\neq 0$, together with Proposition~\ref{preliminaries-on-spectrum}\ref{energy-part}, implies that
\beqa
|\nabla_{\ti r}\nF(q,p,\ti r)|\geq \eps'>0
\eeqa
for some fixed $\eps'$. Thus we can write
\beqa
e^{-i\nF(q,p,\ti r)t}=\fr{\nabla_{\ti r} \nF(q,p,\ti r)\cdot \nabla_{\ti r} e^{-i\nF(q,p,\ti r) t} }{(-it) |\nabla_{\ti r}\nF(q,p,\ti r)|^2}.
\eeqa
Now we define the function
\beqa
J(q,p,\ti r):=\fr{\nabla_{\ti r} \nF(q,p,\ti r)}{ |\nabla_{\ti r}\nF(q,p,\ti r)|^2}  h_1(p-\ti r )h_2(q+\ti r )\chi^{\ka}(\ti r), \label{function-J}
\eeqa
where $\chi^{\ka}\in C_0^{\infty}(\real^3)$ is equal to one on $\mcB_{\ka}$  and vanishes outside of a slightly larger set.
We note that, by Proposition~\ref{preliminaries-on-spectrum}~\ref{energy-part} for any multiindex $\be$ s.t. $0\leq|\be|\leq 1$ 
\beqa
|\pa_{\ti r}^{\be}J(q,p,\ti r)|\leq D(q,p), \label{j-estimate}
\eeqa
where $(q,p)\mapsto D(q,p)$ is a smooth, compactly supported  function. Moreover, for 
 $0\leq|\be|\leq 2$ 
\beqa
|\pa_{\ti r}^{\be} \vv(\ti r)|\leq \fr{\chi_3(\ti r)|\ti r|^{\alf}}{|\ti r|^{\h+|\be|}},\quad
|\pa_{\ti r}^{\be}\chi_j(\ti r)|\leq \fr{c}{(\si_\s)^{|\be|}}, \label{derivative-bounds}
\eeqa
where  $\chi_3$ is a smooth, compactly supported  function, independent of $\si$.
In addition, for  $0\leq|\be|\leq 2$  we obtain from  Theorem~\ref{main-theorem-spectral} \ref{derivatives-bounds} 
\beqa
& &|\pa^{\be}_{\ti r} f^{m}_{p-\ti r,\si}(k)|\leq \fr{1}{\sqrt{m!}} \fr{c}{\si^{\de_{\g_0}} } g^m_{\si}(k),\label{m-derivatives} \\
& &|\pa^{\be}_{\ti r} f^{n+1}_{q+\ti r,\si}(r,\ti r)|\leq \fr{1}{\sqrt{n!}}\fr{c}{\si^{\de_{\g_0}} } \fr{|\ti r|^{\alf}}{|\ti r|^{3/2+|\be|}}  g^{n}_{\si}(r).
\label{n+1-derivatives}
\eeqa

Now using the Gauss Law we obtain from (\ref{decomposition-of-FGH}) 
\beqa
F^{G_1,G_2}_{1,n,m}(q;r \ba p;k)
=\fr{1}{it}\int_{|\ti r|\leq \si_{\s}} d^3\ti r\,  e^{-i\nF(q,p,\ti r)t} \nabla_{\ti r}
 \cdot\bigg(J(q,p,\ti r)  \vv(\ti r) \chi_1(\ti r) f^{n+1}_{q+\ti r ,\si}(r,\ti r )
 f^{m}_{p-\ti r,\si}(k)\bigg)\non\\
 +\fr{\si^2}{it}\int d\Om(\mathbf{n} )\,  e^{-i\nF(q,p,\si n)t} \mathbf{n}
 \cdot\bigg( J(q,p,\ti r)\vv(\ti r) \chi_1(\ti r) f^{n+1}_{q+\ti r ,\si}(r,\ti r ) f^{m}_{p-\ti r,\si}(k)\bigg)\bigg|_{\ti r=\si \mathbf{n}}, \,\,\,
\label{FGH-bound-integration-by-parts}
\eeqa
where $\mathbf{n}$ is the normal vector to the unit sphere and  $d\Om(\mathbf{n}  )$ is the spherical measure (that is 
\beqa
\mathbf{n}=(\sin \theta \cos \vp, \sin \theta \sin \vp, \cos\theta) 
\eeqa
and   $d\Om( \mathbf{n} )=\sin\theta d\theta d\vp$   in  spherical coordinates $(\theta,\phi)$). Let us consider the first term on the r.h.s.
of (\ref{FGH-bound-integration-by-parts}). Let $I_1$ be the corresponding  integrand.
Making use of (\ref{j-estimate})--(\ref{n+1-derivatives}), we obtain
\beqa
|I_1|&\leq&\fr{1}{\sqrt{m!n!}} \sum_{0\leq |\be_1|+|\be_2|+|\be_3|\leq 1}\fr{c}{(\si_\s)^{|\be_1|}}\fr{  D(q,p)}{\si^{\de_{\g_0}} }
\fr{\chi_3(\ti r)|\ti r|^{2\alf}   }{|\ti r|^{2+|\be_2|+|\be_3|}}   g^{n}_{\si}(r)g^m_{\si}(k)\non\\
&\leq&\fr{1}{\sqrt{m!n!}} \fr{c  D(q,p)}{\si^{\de_{\g_0}} }\fr{\chi_3(\ti r)|\ti r|^{2\alf}  }{|\ti r|^{3}}   g^{n}_{\si}(r)g^m_{\si}(k),
\label{I-one-bound}
\eeqa
where in the last step above we made use of the fact that $|\ti r|\leq \si_\s$ in the region of integration.
Now let $I_2$ be the integrand in the boundary integral on the r.h.s. of (\ref{FGH-bound-integration-by-parts}). Making use, again, of 
bounds (\ref{j-estimate})--(\ref{n+1-derivatives}), we get
\beqa
|I_2|&\leq&\fr{1}{\sqrt{m!n!}} \fr{ c D(q,p)}{\si^{\de_{\g_0}} }\fr{\chi_3(\ti r)|\ti r|^{2\alf}   }{|\ti r|^{2} }   g^{n}_{\si}(r)g^m_{\si}(k)
\bigg|_{\ti r=\si\mathbf{n}}\non\\
&=&\fr{1}{\sqrt{m!n!}} \fr{ c D(q,p)}{\si^{\de_{\g_0}} }\fr{\chi_3(\si \mathbf{n}) \si^{2\alf}   }{\si^{2} }   g^{n}_{\si}(r)g^m_{\si}(k).
\label{I-two-bound}
\eeqa
Thus (\ref{FGH-bound-integration-by-parts}), (\ref{I-one-bound}), (\ref{I-two-bound}) give 
\beqa
|F^{G_1,G_2}_{1,n,m}(q;r \ba p;k)|\leq \fr{1}{\sqrt{m!n!}}\fr{ c D(q,p)}{\si^{\de_{\g_0}} } \fr{(\si_\s)^{2\alf} }{t}   g^{n}_{\si}(r)g^m_{\si}(k),
\eeqa
which is the first contribution to the bound (\ref{FGH-bound}).

Now we consider the contribution above  the slow cut-off. In this region we will have to differentiate by parts twice, so a direct
application of the non-stationary phase method would result in third derivatives of the function  $\nF(q,p,\ti r):=E_{ q+\ti r }+E_{p-\ti r }$.
However, our spectral results do not include the existence of third derivatives of $p\mapsto E_p$, but rather the bound
\beqa
|\pa^{\be}_p E_{ p,\si}|\leq \fr{c}{\si^{\de_{\la_0}}} \label{weaker-derivative-bound}
\eeqa
for any multiindex $\be$ s.t. $0\leq |\be|\leq 3$ (cf. Proposition~\ref{preliminaries-on-spectrum} \ref{energy-part}). To be able to exploit
this bound, we introduce an auxiliary function  $\nF_{\si}(q,p,\ti r):=E_{q+\ti r,\si}+E_{p-\ti r,\si}$ and rewrite 
$F^{G_1,G_2}_{2,n,m}(q;r \ba p;k)$, defined in (\ref{decomposition-of-FGH}), as follows
 \beqa
& &F^{G_1,G_2}_{2,n,m}(q;r \ba p;k)
=\int d^3\ti r\, \vv(\ti r) \chi_2(\ti r) e^{-i \nF_{\si}(q,p,\ti r)t}h_1(p-\ti r )h_2(q+\ti r )\cdot f^{n+1}_{q+\ti r ,\si}(r,\ti r ) f^{m}_{p-\ti r,\si}(k)\non\\
& &\pha{4444444444444444444444444444444444444444444444444444444} +R(\si t).
\label{decomposition-of-FGH-one}
\eeqa
The rest term $R(t\si)$ satisfies
\beqa
|R(t\si)|&\leq& \int d^3\ti r\, \vv(\ti r) \chi_2(\ti r) |(1- e^{i(\nF(q,p,\ti r)-\nF_{\si}(q,p,\ti r))t})h_1(p-\ti r )h_2(q+\ti r )
 f^{n+1}_{q+\ti r ,\si}(r,\ti r ) f^{m}_{p-\ti r,\si}(k)|\non\\
&\leq& c\fr{\si t}{ \si^{\de_{\g_0}}} g^m_{\si}(k)    g^{n}_{\si}(r),         
\eeqa
where we made use of the fact that $|\nF(q,p,\ti r)-\nF_{\si}(q,p,\ti r))|\leq c\si$  (Proposition~\ref{preliminaries-on-spectrum}, \ref{energy-part})
and of the bounds (\ref{derivative-bounds}), (\ref{m-derivatives}), (\ref{n+1-derivatives}). This gives the second contribution 
to the bound (\ref{FGH-bound}).

Let us denote by $F^{G_1,G_2}_{2,\si, n,m}$ the first term on the r.h.s. of (\ref{decomposition-of-FGH-one}). We will estimate this term
with the help of the method of non-stationary phase. Similarly as in the first part of the proof
we note that by disjointness of velocity
supports of $h_1$, $h_2$, the condition $h_1(p-\ti r )h_2(q+\ti r )\neq 0$ implies that
\beqa
|\nabla_{\ti r}\nF_{\si}(q,p,\ti r)|\geq \eps''>0 \label{sigma-bound-from-below}
\eeqa
for some fixed $\eps''$, independent of $\si\in (0, \ka]$. (Here we made use of the fact that $S\ni p \mapsto E_{p,\si}$ is strictly
convex, uniformly in $\si$. Cf. Proposition~\ref{preliminaries-on-spectrum} \ref{energy-part}).
 Thus we can write
\beqa
e^{-i\nF_{\si}(q,p,\ti r)t}=\fr{\nabla_{\ti r} \nF_{\si}(q,p,\ti r)\cdot \nabla_{\ti r} e^{-i\nF_{\si}(q,p,\ti r) t} }{(-it) |\nabla_{\ti r}\nF_{\si}(q,p,\ti r)|^2}.
\eeqa
We define the function
\beqa
J_{\si}(q,p,\ti r):=\fr{\nabla_{\ti r} \nF_{\si}(q,p,\ti r)}{ |\nabla_{\ti r}\nF_{\si}(q,p,\ti r)|^2}  h_1(p-\ti r )h_2(q+\ti r )\chi^{\ka}(\ti r), \label{function-J-si}
\eeqa
which is analogous to the function $J$ introduced in (\ref{function-J}) above. Making use of (\ref{weaker-derivative-bound})
and of (\ref{sigma-bound-from-below}), we get for all multiindices $\be$ s.t. $0\leq |\be|\leq 2$
\beqa
|\pa_{\ti r}^{\be} J_{\si}(q,p,\ti r)|\leq \fr{D(q,p)}{\si^{\de_{\la_0}}  }.  \label{J-sigma-bound}
\eeqa
By integrating  twice by parts in the defining expression for $F^{G_1,G_2}_{2,\si, n,m}$, we get
\beqa
F^{G_1,G_2}_{2,\si, n,m}(q; r \ba p; k)=\fr{1}{(it)^2}\int d^3\ti r\,  e^{-i\nF_{\si}(q,p,\ti r)t}\cdot\pha{4444444444444444444444444444}\non\\
\times\nabla_{\ti r} \cdot\bigg( \fr{\nabla_{\ti r}\nF_{\si}(q,p,\ti r)}{ |\nabla_{\ti r}\nF_{\si}(q,p,\ti r)|^2}\nabla_{\ti r} 
\cdot\big( J_{\si}(q,p,\ti r)  \vv(\ti r) \chi_2(\ti r) f^{n+1}_{q+\ti r ,\si}(r,\ti r )
 f^{m}_{p-\ti r,\si}(k)\big)\bigg). \label{second-derivative}
\eeqa
In view  of (\ref{weaker-derivative-bound}), (\ref{sigma-bound-from-below}), the function 
\beqa
(q,p,\ti r)\mapsto \fr{\nabla_{\ti r}\nF_{\si}(q,p,\ti r)}{ |\nabla_{\ti r}\nF_{\si}(q,p,\ti r)|^2}
\eeqa
is bounded by $c/\si^{\de_{\la_0}}$, together with its first  derivatives, on the support of 
$(q,p,\ti r)\mapsto h_1(p-\ti r )h_2(q+\ti r )\chi^{\ka}(\ti r)$. Thus we obtain from the bounds (\ref{derivative-bounds}), (\ref{m-derivatives}), 
(\ref{n+1-derivatives}) and (\ref{J-sigma-bound}) that the integrand $I$ in (\ref{second-derivative})
satisfies
\beqa
|I|&\leq& \fr{1}{\sqrt{m!n!}}\sum_{0\leq |\be_1|+|\be_2|+|\be_3|\leq 2}\fr{c}{(\si_\s)^{|\be_1|}}\fr{  D(q,p)}{\si^{\de_{\g_0}} }
\fr{\chi_3(\ti r)|\ti r|^{2\alf}   }{|\ti r|^{2+|\be_2|+|\be_3|}}   g^{n}_{\si}(r)g^m_{\si}(k)\non\\
&\leq& \fr{1}{\sqrt{m!n!}}   \fr{c}{(\si_\s)^{2}}\fr{  D(q,p)}{\si^{\de_{\g_0}} }
\fr{\chi_3(\ti r)|\ti r|^{2\alf}   } {|\ti r|^{2}}   g^{n}_{\si}(r)g^m_{\si}(k), 
\eeqa
where in the second step we made use of the fact that $(1-\eps)\si_\s\leq |\ti r|$ in the region of integration.
Thus we get from (\ref{second-derivative}) that
\beqa
|F^{G_1,G_2}_{2,\si, n,m}(q; r\ba p; k)|\leq \fr{1}{\sqrt{m!n!}}\fr{ c D(q,p)}{\si^{\de_{\g_0}} } \fr{1 }{t^2 (\si_\s)^2}   g^{n}_{\si}(r)g^m_{\si}(k),
\eeqa
which gives the third contribution to (\ref{FGH-bound}). The factor $(n+1)$, appearing in (\ref{factor-n+1}), can be estimated by $2^n$
and incorporated into the constant appearing in the definition of $g^{n}_{\si}$. \qed

\bel\label{second-contribution-decay} Let $\check I^{(1)}_{m,n,\ti m,\ti n}$ be defined as follows 
\beqa
\check I^{(1)}_{m,n,\ti m,\ti n}&:=&\int  d^3 q d^3p \int  d^{3n}r d^{3m}k\,  
G_{1,n}(q;  r) G_{2,m}(p;  k)\non\\
& &\pha{4444444444444444444}\times\ov G_{1,\ti n}'( q+\un{\check{k}}-\un{\check{r}}; \hat r, \check k)
\ov  G_{2,\ti m}'(p-\un{\check{k}}+\un{\check{r}}; \hat k, \check r),
\eeqa
where $G_{1,n}$, $G_{2,\ti n}'$ appeared in (\ref{G-def-repeated}), (\ref{G-prime-def}), 
the notation
$k=(\hat k, \check k)$, $r=(\hat r, \check r)$ is explained in Lemma~\ref{combinatorics} and we consider a permutation
in (\ref{B-expectation-value}) for which  $\check k$
or $\check r$ are non-empty. Then there holds
\beqa
|\check I^{(1)}_{m,n,\ti m,\ti n}|\leq  \bigg(\fr{1}{t}\bigg(\fr{1}{\si^{\de_{\g_0}}}+\fr{1}{\si^{ 1/(8\tiga) } }\bigg)+(\sip)^{2\alf} \bigg)
 \fr{ C(h,h') }{\sqrt{m!n!\ti m! \ti n!} } ( \|g^{n-1}_{\si}\|^2_2 +\|g^n_{\si}\|^2_2) (\|g^{m-1}_{\si}\|^2_2+\|g^m_{\si}\|_2^2 ),\,\,\,   \label{I-two-final-formula}
\eeqa
where $C(h,h'):=c\|h_1\|_1\|h_2\|_1 \sum_{\be_1,\be_2; 0\leq |\be_1|+|\be_2|\leq 1} \|\pa^{\be_1}h_1'\|_{\infty} \|\pa^{\be_2}h_2'\|_{\infty}$
and the sum extends over multiindices $\be_1$, $\be_2$. We note that for $\check k$ (resp. $\check r$) non-empty we have $m\neq 0$ (resp. $n\neq 0$)
 and there always holds $m+n=\ti m+\ti n$. We set by convention $g^{-1}_{\si}=0$.
\eel
\proof By inserting the definitions of $G_{i,n}$, $G_{i,\ti n}'$,  $i\in\{ 1,2 \}$, 
we obtain
\beqa
\check I^{(1)}_{m,n,\ti m,\ti n}
&=&\int  d^3 q d^3p \int  d^{3n}r d^{3m}k\, 
e^{i(E_{ p-\un{\check k}+\un{\check r}     }-E_{q})t}e^{ i(E_{q+\un{\check{k}}-\un{\check{r}}    }-E_p)t} 
  h_1(p)h_2(q)    \non\\
& &\times \ov h_1'(  p-\un{\check k}+\un{\check r}       )  \ov h_2'(  q+\un{\check{k}}-\un{\check{r}}     ) f^{n}_{q,\si}(r)  
f^{m}_{p,\si}(k)\ov f^{\ti n}_{  q+\un{\check{k}}-\un{\check{r}}     ,\sip}(\hat r, \check k)
\ov f^{\ti m}_{   p-\un{\check k}+\un{\check r}    ,\sip}( \hat k, \check r).\label{I-two-formula}
\eeqa 
Since the expression on the r.h.s. of (\ref{I-two-formula}) and the bound (\ref{I-two-final-formula}) are invariant under the substitutions $k\leftrightarrow r$, $p\leftrightarrow q$,  $h_1\leftrightarrow h_2$, $h'_1\leftrightarrow h'_2$, $m\leftrightarrow n$, $\ti m \leftrightarrow \ti n$ it suffices to consider the case of non-empty $\check k$.
Similarly as in the proof of Lemma~\ref{smooth-bounds}, we introduce a slow infrared  cut-off.
Let us set $\sip_\s:=\ka (\sip/\ka)^{1/(8\tiga)}$, which clearly satisfies $\sip\leq\sip_\s\leq \ka$.
Let $\chi\in C^{\infty}(\real^3)$, $0\leq \chi\leq 1$, be supported in $\mcB_1$ (the unit ball) and be equal to one on $\mcB_{1-\eps}$ for
some  $0<\eps<1$. We set $\chi_1(\chk_1  ):=\chi( \chk_1/\sip_\s)$, $\chi_2( \chk_1  ):= 1-\chi_1( \chk_1)$, where $\chk_1$
is the first component of $\chk$, and 
write for $j\in\{1,2\}$
\beqa
\check I^{(1)(j)}_{m,n,\ti m,\ti n}
&:=&\int  d^3 q d^3p \int  d^{3n}r d^{3m}k\, 
e^{i(E_{ p-\un{\check k}+\un{\check r}     }-E_{q})t}e^{ i(E_{q+\un{\check{k}}-\un{\check{r}}    }-E_p)t} \non\\
& & \times h_1(p)h_2(q)  \ov h_1'(  p-\un{\check k}+\un{\check r}       ) \ov  h_2'(  q+\un{\check{k}}-\un{\check{r}}     ) 
 f^{n}_{q,\si}(r) \,
 f^{m}_{p,\si}(k) \,\non\\
& &\times\big( \chi_j(\chk_1) \ov f^{\ti n}_{  q+\un{\check{k}}-\un{\check{r}}     ,\sip}(\hat r, \check k)\big) \,
\ov f^{\ti m}_{   p-\un{\check k}+\un{\check r}    ,\sip}( \hat k, \check r). \label{I-two-formula-two}
\eeqa
Let us first consider (\ref{I-two-formula-two}) with $j=1$. We conclude  from Theorem~\ref{main-theorem-spectral} and the definition
of the functions $g^n_{\si}$  that
\beqa
& &|\chi_1(\chk_1)f^{n}_{q,\si}(r) f^{m}_{p,\si}(k) f^{\ti n}_{q+\un{\check{k}}-\un{\check{r}}     ,\sip}(\hat r, \check k) 
 f^{\ti m}_{   p-\un{\check k}+\un{\check r}    ,\sip}( \hat k, \check r)|\non\\
& &\pha{44444}\leq\fr{1}{\sqrt{m!n!\ti m!\ti n!}}\chi_1(\chk_1)g^n_{\si}(r) g^{m}_{\si}(k)  g^{\ti n}_{\sip}(\hat r, \check k)g^{\ti m}_{\sip}(\hat k, \check r) \non\\
& &\pha{44444}\leq \fr{1}{\sqrt{m!n!\ti m!\ti n!}}\fr{c\chi(\chk_1/\sip_\s)\chi_{[\sip,\kas)}(\chk_1)^2 |\chk_1|^{2\alf}    }{|\chk_1|^{3}} 
g^n_{\si}(r) g^{m-1}_{\si}(k')  g^{\ti n-1}_{\sip}(\hat r, \chk')g^{\ti m}_{\sip}(\hat k, \check r)\non\\
& &\pha{44444}\leq \fr{1}{\sqrt{m!n!\ti m!\ti n!}}\fr{c\chi(\chk_1/\sip_\s)\chi_{[\sip,\kas)}(\chk_1)^2  |\chk_1|^{2\alf}    }{|\chk_1|^{3}}   g^n_{\si}(r)^2 g^{m-1}_{\si}(k')^2, \label{pointwise-bounds}
\eeqa
where we decomposed $k=(\chk_1,k')$, $\chk=(\chk_1,\chk')$  and in the last step we made use of the fact that $g^{\ti m}_{\sip}(\hat k,\check r)\leq g^{\ti m}_{\si}(\hat k,\check r)$ and $g^{\ti n-1}_{\sip}(\hat r, \chk')\leq g^{\ti n-1}_{\si}(\hat r, \chk')$  for $\sip\geq \si$. Substituting~(\ref{pointwise-bounds}) to 
(\ref{I-two-formula-two}), we get 
\beqa
|\check I^{(1)(1)}_{m,n,\ti m,\ti n}|\leq (\sip_\s)^{2\alf} \fr{c\|h_1\|_1\|h_2\|_1\|h_1'\|_{\infty} \|h_2'\|_{\infty} }{\sqrt{m!n!\ti m!\ti n!}} \|g^n_{\si}\|^2_2 \|g^{m-1}_{\si}\|^2_2.
\label{below-slow-cut-off-part}
\eeqa

Let us now consider (\ref{I-two-formula-two})  for $j=2$. We  set $w:=\un{\check{k'}}-\un{\check{r}}$ and define  
\beqa
\nF(\chk_1, p,q,w):=E_{-\chk_1+p-w}+E_{\chk_1+q+w}. 
\eeqa
We note that, by disjointness of velocity supports of $h_1$, $h_2$,  the condition $h_1(-\chk_1+p-w )h_2(\chk_1+q+w)\neq 0$ implies that
\beqa
|\nabla_{\chk_1}  \nF( \chk_1, p, q, w)|\geq\eps'>0, \label{two-gradient-positivity}
\eeqa
for some $\eps'$ independent of $\chk_1$, $p$, $q$ and $w$. Thus we can write the following identity
\beqa
e^{i  \nF( \chk_1, p,q,w  ) t }=\fr{\nabla_{\chk_1} \nF(\chk_1, p,q,w   )\cdot \nabla_{\chk_1} e^{i  \nF(\chk_1, p,q,w )t} }{ i t|\nabla_{ \chk_1} \nF(\chk_1, p,q,w) |^2 }.
\eeqa
Now we define the function
\beqa
J(\chk_1, p,q,w ):=\fr{\nabla_{\chk_1}  \nF( \chk_1, p,q,w  ) }{ |\nabla_{\chk_1}  \nF( \chk_1, p,q,w  )|^2}  \ov h_1'(-\chk_1+p-w )\ov h_2'(\chk_1+q+w). 
\eeqa
We note that, by Proposition~\ref{preliminaries-on-spectrum} \ref{energy-part}, for any multiindex $\be$ s.t. $ 0\leq|\be|\leq 1$ there hold the bounds
\beqa
|\pa_{ \chk_1}^{\be} J(\chk_1, p,q,w)|\leq c_0(h_1',h_2'),\quad
|\pa_{  \chk_1 }^{\be}\chi_j( \chk_1  )|\leq \fr{c}{(\sip_\s)^{|\be|} },\label{derivative-bounds-one}
\eeqa
where $c_0(h_1',h_2')$ has the form
\beqa
c_0(h_1',h_2')=c\sum_{\be_1,\be_2; 0\leq |\be_1|+|\be_2|\leq 1} \|\pa^{\be_1}h_1'\|_{\infty} \|\pa^{\be_2}h_2'\|_{\infty},
\eeqa 
where $\be_1,\be_2$ are multiindices. Moreover, we obtain from  Theorem~\ref{main-theorem-spectral} \ref{derivatives-bounds} 
\beqa
|f^{n}_{q,\si}(r) |&\leq& \fr{1}{\sqrt{n!}}g^n_{\si}(r),\label{no-der-estimate}\\
|\pa_{\chk_1}^{\be} f^{\ti m}_{   p-\un{\check k}+\un{\check r}    ,\sip}( \hat k, \check r)|&\leq& \fr{1}{\sqrt{\ti m!}}
\bigg(\fr{1}{(\sip)^{\de_{\la_0}}  }\bigg)^{|\be|} g^{\ti m}_{\sip}( \hat k, \check r ), \label{der-estimate-one} \\
|\pa_{\chk_1}^{\be} \big(\chi_2(\check k_1)\ov f^{\ti n}_{  q+\un{\check{k}}-\un{\check{r}}     ,\sip}(\hat r, \check k)f^{m}_{p,\si}(k)\big)|&\leq &\fr{1}{\sqrt{\ti n!m!}}
\bigg(\fr{1}{(\sip)^{\de_{\la_0}}  }+\fr{1}{\sip_\s}\bigg)^{|\be|}  g^{\ti n}_{\sip}(\hat r, \check k)g^m_{\si}(k).\quad\quad \label{der-estimate-two}
\eeqa
Now coming back to formula~(\ref{I-two-formula-two}) and integrating by parts we obtain
\beqa
\check I^{(1)(2)}_{m,n,\ti m,\ti n}
&=&-\fr{1}{it}\int  d^3 q d^3p \int  d^{3n}r d^{3m}k\, 
e^{i( \nF( \chk_1, p,q,w  )- \nF( 0,p,q,0 ))t}  h_1(p)h_2(q)  f^{n}_{q,\si}(r)   \,\non\\
& & \times \nabla_{\chk_1}\cdot\bigg( J(\chk_1, p,q,w ) 
\chi_2(\check k_1)\ov f^{\ti n}_{  q+\un{\check{k}}-\un{\check{r}}     ,\sip}(\hat r, \check k) \,
f^{m}_{p,\si}(k)\ov f^{\ti m}_{   p-\un{\check k}+\un{\check r}    ,\sip}( \hat k, \check r)\bigg).\label{I-two-non-stationary}
\eeqa
Making use of the bounds (\ref{derivative-bounds-one}), (\ref{no-der-estimate}), (\ref{der-estimate-one}), (\ref{der-estimate-two}) and
of the fact that $g^{\ti n}_{\sip}(\hat r, \check k)\leq g^{\ti n}_{\si}(\hat r, \check k)$, since $\sip\geq \si$, we estimate
\beqa
|\check I^{(1)(2)}_{m,n,\ti m,\ti n}|&\leq& \fr{1}{t}\bigg(\fr{1}{(\sip)^{\de_{\la_0}}  }+\fr{1}{\sip_\s}\bigg)
 \fr{\|h_1\|_1\|h_2\|_1c_0(h_1',h_2')}{\sqrt{m!n!\ti m!\ti n!}}  \| g^n_{\si}\|_2^2   \|g^m_{\si}\|_2^2. \label{above-slow-cut-off-part}
\eeqa
Exploiting (\ref{above-slow-cut-off-part}), (\ref{below-slow-cut-off-part}) and  the fact that $\sip_\s=\ka(\sip/\ka)^{1/(8\tiga)}$, 
we get (\ref{I-two-final-formula}). Here we estimated trivially
\beqa
 \|g^n_{\si}\|^2_2 \|g^{m-1}_{\si}\|_2^2+\|g^n_{\si}\|^2_2 \|g^m_{\si}\|_2^2 \leq  ( \|g^{n-1}_{\si}\|^2_2 +\|g^n_{\si}\|^2_2) (\|g^{m-1}_{\si}\|^2_2+\|g^m_{\si}\|_2^2 )
\eeqa
to obtain an expression in (\ref{I-two-final-formula}) which is symmetric under the substitution $m\leftrightarrow n$. \qed

\bel\label{First-contribution-decay} Let $\hat I^{(2)}_{m,n,\ti m,\ti n}$ be defined as follows 
\beqa
\hat I^{(2)}_{m,n,\ti m,\ti n}&:=&\int  d^3 q d^3p \int  d^{3n}r d^{3m}k\,  
G_{1,n}(q;  r) G_{2,m}(p;  k)\non\\
& &\pha{444444444444444444}\times\ov G_{1,\ti n}'(p-\un{\hat k}+\un{\hat r}; \hat r, \check k) \ov G_{2,\ti m}'(q+\un{\hat{k}}-\un{\hat{r}}; \hat k, \check r),
\eeqa
where $G_{i,m}$, $G_{i, m}'$, $i\in\{1,2\}$, are defined in (\ref{G-def-repeated}), (\ref{G-prime-def}), the notation
$k=(\hat k, \check k)$, $r=(\hat r, \check r)$ is explained in Lemma~\ref{combinatorics}  and we consider a permutation
in (\ref{B-expectation-value}) for which  $\hat k$
or $\hat r$ are non-empty. Then there holds
\beqa
|\hat I^{(2)}_{m,n,\ti m,\ti n}|\leq  \bigg(\fr{1}{t}\bigg(\fr{1}{\si^{\de_{\g_0}}}+\fr{1}{\si^{ 1/(8\tiga) } }\bigg)+(\sip)^{2\alf} \bigg)
 \fr{C(h,h')}{\sqrt{m!n!\ti m! \ti n!} } ( \|g^{n-1}_{\si}\|^2_2 +\|g^n_{\si}\|^2_2) (\|g^{m-1}_{\si}\|^2_2+\|g^m_{\si}\|_2^2 ),  \,\,\,\,  \label{I-one-final-formula}
\eeqa
where $C(h,h'):=c\|h_1\|_1\|h_2\|_1 \sum_{\be_1,\be_2; 0\leq |\be_1|+|\be_2|\leq 1} \|\pa^{\be_1}h_1'\|_{\infty} \|\pa^{\be_2}h_2'\|_{\infty}$
and the sum extends over multiindices $\be_1$, $\be_2$.  We note that for $\hat k$ (resp. $\hat r$) non-empty we have $m\neq 0$ (resp. $n\neq 0$)
 and there always holds $m+n=\ti m+\ti n$. We set by convention $g^{-1}_{\si}=0$.
\eel
\proof By inserting the definitions of $G_{i,m}$, $G_{i, m}'$ we obtain
\beqa
\hat I^{(2)}_{m,n,\ti m,\ti n}
=\int  d^3 q d^3p \int  d^{3n}r d^{3m}k\, 
e^{i(E_{q+\un{\hat{k}}-\un{\hat{r}} }-E_q)t}e^{ i(E_{p-\un{\hat k}+\un{\hat r}}-E_p)t} 
  h_1(p)h_2(q)   \non\\
\times \ov h_1'(q+\un{\hat{k}}-\un{\hat{r}} ) \ov  h_2'(p-\un{\hat k}+\un{\hat r}  ) f^{n}_{q,\si}(r)  
\ov f^{\ti m}_{q+\un{\hat{k}}-\un{\hat{r}} ,\sip}( \hat k, \check r) f^{m}_{p,\si}(k)
\ov f^{\ti n}_{ p-\un{\hat k}+\un{\hat r} ,\sip}(\hat r, \check k).
\label{I-one-formula}
\eeqa 
We note that by substitutions $ h_1'\leftrightarrow h_2'$, $\ti m\leftrightarrow \ti n$, $\hk \leftrightarrow \chk$, $\hr\leftrightarrow \chr$
we obtain formula~(\ref{I-two-formula}). Now the statement follows from Lemma~\ref{second-contribution-decay} and the fact
that (\ref{I-two-final-formula}) does not change under the above substitutions. \qed

\appendix

\section{Domain questions} \label{domain-questions}
\setcounter{equation}{0}

\bel\label{self-adjointness} 
There exist constants $0\leq a<1$  and $b\geq 0$ s.t. for any  $\Psi\in \mcC^{(n)}$ there holds the bound
\beqa
\|H_{\I}^{(n)}\Psi\|\leq a\|H_{\free}^{(n)}\Psi\|+b\|\Psi\|, \label{Kato-Relich}
\eeqa 
where $H_{\I}^{(n)}=H_{\I}|_{\mcC^{(n)}}$, $H_{\free}^{(n)}=H_{\free}|_{\mcC^{(n)}}$ and the constant $b$ may depend on $n$.
\eel
\proof Let us use the form of the interaction Hamiltonian appearing in formula (\ref{explicit-Hamiltonian}). We have
 $H_{\I}^{(n)}=H_{\I}^{\ain, (n)}+H_{\I}^{\cin, (n)}$, where 
\beqa
H_{\I}^{\ain, (n)}:=\sum_{i=1}^n\int d^3k\, \vv(k)\, e^{ikx_i}a(k)
\eeqa
and $H_{\I}^{\cin, (n)}=(H_{\I}^{\ain, (n)})^*$. 
Let us set $C_n(k):= \sum_{l=1}^n e^{ikx_l}$ and compute for some $\Psi\in \mcC^{(n)}$
\beqa
\|H_{\I}^{\ain,(n)}\Psi\|&\leq& \int d^3k\,\vv(k)\|C_n(k) a(k)\Psi\|\leq n \int d^3k\,\vv(k)\|a(k)\Psi\|\non\\
&\leq& n\|\om^{-1/2}\vv\|_2\lan \Psi,H_{\pho}\Psi\ran^{\h}\leq \fr{1}{4}\|H_{\free}\Psi\|+n^2\|\om^{-1/2}\vv\|_2^2\|\Psi\|, \label{KR-bound}
\eeqa
where in the last step we anticipate that  (\ref{Kato-Relich}) should hold with $0<a<1$.

Let us now consider the creation part of $H_{\I}^{(n)}$. Making use of the canonical commutation relations, we get
\beqa
\|H_{\I}^{\cin,(n)}\Psi\|^2&=&\int d^3k_1d^3k_2\,\vv(k_1)\vv(k_2)\lan C_n(k_1)^*a^*(k_1)\Psi,C_n(k_2)^*a^*(k_2)\Psi\ran\non\\
&\leq &\|H_{\I}^{\ain,(n)}\Psi\|^2+  n^2 \|\vv\|_2^2   \|\Psi\|^2,
\eeqa
which, together with (\ref{KR-bound}), concludes the proof. \qed
\bel \label{ren-creation-operator} The domain $\mcD$, defined in (\ref{invariant-domain}), is contained in the
domains of  $H, H_{\el}, H_{\pho}, H_{\I}^{\ain/\cin}, \cH_{\I,\si}$ and $\nr^*(h)$, $h\in \B(\PS)$. Moreover, these
operators leave $\mcD$ invariant.
\eel
\proof Let $\Psi_l^{r_1,r_2}$ be a vector of the form (\ref{psi-l}). Then
\beqa
& &\nr^*_{\si}(h)\Psi_l^{r_1,r_2}\non\\
&=&\sum_{m,m'}\fr{1}{\sqrt{m'!}} \fr{1}{\sqrt{m!}} \int d^3p'\,  d^{3l}p\,d^{3m'}k'\, d^{3m}k\, 
F'_{1,m'}(p';k')F_{l,m}(p;k)   \non\\
& &\ph{44444444444444444444444444}\times \nn^*(p') \nn^*(p)^l \bb^*(k')^{m'}\bb^*(k)^m\vac \non\\
&=& \sum_{\ti m=0}^{\infty}\fr{1}{\sqrt{\ti m!}}    
\int   d^{3(l+1)}\ti p\, d^{3\ti m}\ti k \, \ti F_{\ti m}(\ti p;\ti k)  \nn^*(\ti p)^{l+1} \bb^*(\ti k)^{\ti m}\vac,    
\eeqa
where $F'_{1,m'}(p',k'):= h(p'+\unk') f^{m'}_{p'+\unk,\si}(k')$, $\ti m:=m+m'$, $\ti k:=(k,k')$, $\ti p:=(p,p')$ and
\beqa
\ti F_{\ti m}(\ti p;\ti k)=\sum_{m=0}^{\ti m} \fr{ \sqrt{\ti m!} }{   \sqrt{(\ti m-m)!} \sqrt{m!}  }(F'_{1,(\ti m-m) }F_{l,m})_{\sym}  (\ti p;\ti k),
\eeqa
where the symmetrization is performed in the $\ti p$ and $\ti k$ variables separately. By Theorem~\ref{main-theorem-spectral},   $F'_{1,m'}$
satisfies the bound  (\ref{factorial-bound}), and therefore
\beqa
\|\ti F_{\ti m}\|_2\leq \fr{c^{\ti m}}{\sqrt{\ti m!}},
\eeqa
for some constant $c$. Hence $\nr^*_{\si}(h)\Psi_l^{r_1,r_2}$ is well defined and belongs to $\mcD$. 

Next, we note that
\beqa
H_{\el}\Psi_l^{r_1,r_2}=\sum_{m=0}^{\infty}\fr{1}{\sqrt{m!}}\int d^{3l}p d^{3m}k\,(\Om(p_1)+\cdots+\Om(p_l) ) F_{l,m}(p;k)\nn^*(p)^l\bb^*(k)^m\vac,\\
H_{\pho}\Psi_l^{r_1,r_2}=\sum_{m=0}^{\infty}\fr{1}{\sqrt{m!}}\int d^{3l}p d^{3m}k\,(\om(k_1)+\cdots+\om(k_m) ) F_{l,m}(p;k)\nn^*(p)^l\bb^*(k)^m\vac.
\eeqa
Due to the support properties of $F_{l,m}(p;k)$ these vectors are well defined and belong to $\mcD$.

Finally, we consider the operators  $H_{\I}^{\ain/\cin}$. We recall that the interaction Hamiltonian restricted to $\hil^{(l)}$
has the form
\beqa
H_{\I}^{(l)}&:=&\sum_{i=1}^l\int d^3k\, \vv(k)\, ( e^{ikx_i}a(k)+e^{-ikx_i}a^*(k) )
\eeqa
and we express $\Psi_l^{r_1,r_2}$ in terms of its $m$-photon components, i.e., 
\beqa
\{\Psi_l^{r_1,r_2}\}^{(l,m)}(p_1,\ldots, p_l;k_1,\ldots,k_m)=F_{l,m}(p_1,\ldots, p_l;k_1,\ldots, k_m),
\eeqa
Now we can write
\beqa
(H_{\I}^{\ain}\Psi_l^{r_1,r_2})^{(l,m)}(p_1,\ldots, p_l; k_1,\ldots, k_m)\ph{4444444444444444444444444444444}\non\\
=\sqrt{m+1}\int d^3k\,\vv(k)\sum_{i=1}^l (\Psi_l^{r_1,r_2})^{(l,m+1)}(p_1,\ldots, p_i-k,\ldots, p_l; k, k_1,\ldots, k_m).
\eeqa
It is easy to see that for some constant $c$, independent of $m$,
\beqa
\|(H_{\I}^{\ain}\Psi_l^{r_1,r_2})^{(l,m)}\|_2\leq \fr{c^m}{\sqrt{m!}}. \label{c-m-bound}
\eeqa
Similarly, we obtain that
\beqa
(H_{\I}^{\cin}\Psi_l^{r_1,r_2})^{(l,m)}(p_1,\ldots, p_l; k_1,\ldots, k_m)\ph{4444444444444444444444444444444}\non\\
=\fr{1}{\sqrt{m}}\sum_{i=1}^m\sum_{j=1}^l\vv(k_i)(\Psi_l^{r_1,r_2})^{(l,m-1)}(p_1,\ldots,p_j+k_i,\ldots, p_l; k_1,\ldots,k_{i_*},\ldots, k_m),
\eeqa
where $k_{i_*}$ means omission of the $i$-th variable.
This gives, again, a bound of the form (\ref{c-m-bound}). Since the case of $\cH_{\I,\si}$ is analogous, this concludes the proof. \qed

\section{Fock space combinatorics}\label{Fock-space-combinatorics}
\setcounter{equation}{0}

In Lemmas~\ref{norms-of-single-particle-states}, \ref{double-creation-lemma} and  \ref{norms-of-scattering-states-one}
below we verify the identities first for $G_{i,m}, G_{i,m}', F_{n,m}, F'_{n,m}$ of Schwartz class and then extend them to
square integrable functions using Theorem X.44 of \cite{RS2}
\bel\label{norms-of-single-particle-states} Let  $G_m, G_m'\in L^2(\real^3\times \real^{3m})$ be symmetric in their photon variables, 
see (\ref{notation-G-m}).
Let us define as operators on $\mcC$
\beqa
B_m^*(G_m):=\int d^3p\,d^{3m}k\, G_m(p;k) \neta^*(p-\unk)a^*(k)^m
\eeqa
and $B_m(G_m):=(B_m^*(G_m))^*$.
Then there holds the identity
\beqa
\lan\vac, B_m(G_m') B_m^*(G_m) \vac\ran=m!\int d^3p  d^{3m}k\,  \ov  G_{m}'(p;  k) G_{m}(p;  k ) .
\eeqa
\eel
\proof We compute
\beqa
& &\lan\vac, B_m(G_m') B_m^*(G_m) \vac\ran\non\\
&=&\int d^3p\,d^{3m}k\,\int d^3p'\,d^{3m}k'\, G_m(p;k)  \ov G_{m}'(p'; k')                    
 \lan\vac, a(k')^m\neta(p'-\unk') \neta^*(p-\unk)a^*(k)^m\vac\ran\non\\
&=&\int d^3p\,d^{3m}k\,\int d^3p'\,d^{3m}k'\, G_m(p;k)  \ov G_{m}'(p'; k')                    
\de(p-\unk-p'+\unk') \lan\vac, a(k')^m a^*(k)^m\vac\ran\non\\
&=&\int d^3p\,d^{3m}k\,\int\,d^{3m}k'\, G_m(p;k)  \ov G_{m}'(p-\unk+\unk'; k')                    
 \lan\vac, a(k')^m a^*(k)^m\vac\ran\non\\
&=& \int d^3p\,d^{3m}k\,\int\,d^{3m}k'\, G_m(p;k)  \ov G_{m}'(p-\unk+\unk'; k')                    
 \sum_{\rho\in S_m}\prod_{i=1}^m\de(k_{\rho(i)}-k'_i) \non\\
&=& m!\int d^3p\,d^{3m}k\, G_m(p;k)  \ov G_{m}'(p; k),                    
 \eeqa
where $S_m$ is the set of all permutations of an $m$-element set and in the last step we exploited the fact that $G_{m}'$
is symmetric in its photon variables. \qed
\bel\label{combinatorics} Let $n,m,\ti n,\ti m\in \nat_0$ be s.t. $n+m=\ti n+\ti m$.  Let us choose
\beqa
 r=(r_1,\ldots,r_n)\in \real^{3n},& & 
k=(k_1,\ldots, k_m)\in\real^{3m},  \\
\ti r=(\ti r_1,\ldots, \ti r_{\ti n})\in \real^{3\ti n}, & &
\ti k=(\ti k_1,\ldots,\ti k_{\ti m}  )\in\real^{3\ti m}
\eeqa
and define the sets
\beqa
& &C_n:=\{1,\ldots, n\}, \quad C_n':=\{n+1,\ldots,n+m\}, \\
& &C_{\ti n}:=\{1,\ldots,\ti n\}, \quad  C_{\ti n}':=\{\ti n+1,\ldots,\ti n+ \ti m\}. 
\eeqa
(Note that $C_n'$ is the complement of $C_n$ in $\{1,\ldots, n+m\}$. Similarly for $C_{\ti n}'$).
Let $S_{m+n}$ be the set of permutations of an $m+n$ element set. 
For any $\sig\in S_{m+n}$ we introduce the following notation: 
\beqa
& &\hr:=(r_i)_{(i,\sig(i))\in C_n\times C_{\ti n}},\quad\quad \chr:=(r_i)_{(i,\sig(i))\in C_n\times C_{\ti n}' },\\
& &\hk:=(k_{i-n})_{(i,\sig(i))\in C_n'\times C_{\ti n}' },\quad \chk:=(k_{i-n})_{(i,\sig(i))\in C_n'\times C_{\ti n} },
\eeqa
so that $r=(\hr,\chr)$, $k=(\hk,\chk)$. Similarly,
\beqa
& &\hat{\ti r}:=(\ti{r}_{\sig(i)})_{(i,\sig(i))\in C_n\times C_{\ti n}},\quad\quad \check{\ti r}:=(\ti r_{\sig(i)})_{(i,\sig(i))\in C_n'\times C_{\ti n} },\\
& &\hat{\ti k}:=(\ti k_{\sig(i)-\ti n})_{(i,\sig(i))\in C_n'\times C_{\ti n}' },\quad \check{\ti k}:=(\ti k_{\sig(i)-\ti n })_{(i,\sig(i))\in C_n\times C_{\ti n}' },
\eeqa
so that $\ti r=(\hat{\ti r},  \check{\ti r})$ and $\ti k=(\hat{\ti k}, \check{\ti k})$. (If $\{\, i\,|\, (i,\sig(i))\in C_n\times C_{\ti n}\,\}=\emptyset$
then we say that $\hr$ is empty, and analogously for other collections of photon variables introduced above).
Finally, we define
\beqa
\de(\hr-\hat{\ti  r})&:=&\prod_{(i,\sig(i))\in C_n\times C_{\ti n} } \de(r_i-\ti r_{\sig(i)}),\\
\de(\chr-\check{\ti k})&:=&\prod_{(i,\sig(i))\in C_n\times C_{\ti n}' } \de(r_i-\ti k_{\sig(i)-\ti n} ),\\
\de(\chk-\check{\ti r})&:=&\prod_{(i,\sig(i))\in C_n'\times C_{\ti n} } \de(k_{i-n}-\ti r_{\sig(i)} ), \\
\de(\hk-\hat{\ti k} )&:=&\prod_{(i,\sig(i))\in C_n'\times C_{\ti n}' } \de(k_{i-n}-\ti k_{\sig(i)-\ti n} ).
\eeqa 
Then there holds
\beqa
\lan\vac, \bb(\ti r)^{\ti n} \bb(\ti k)^{\ti m} \bb^*(r)^{n} \bb^*(k)^{m}\vac\ran=\sum_{\sig\in S_{m+n}}
\de(\hr-\hat{\ti  r})\de(\chr-\check{\ti k})\de(\chk-\check{\ti r})\de(\hk-\hat{\ti k} ).
\eeqa
\eel
\proof Let $(v_1,\ldots, v_{n+m})=(r_1,\ldots,r_n,k_1,\ldots, k_m)$
and $(\ti v_1,\ldots,\ti v_{n+m})=(\ti r_1,\ldots,\ti r_{\ti n},\ti k_1,\ldots, \ti k_{\ti m})$. There holds
\beqa
& &\lan\vac, \bb(\ti r)^{\ti n} \bb(\ti k)^{\ti m} \bb^*(r)^{n} \bb^*(k)^{m}\vac\ran=\sum_{\sig\in S_{m+n}}\prod_{j=1}^{m+n}\de(v_{j}-\ti v_{\sig(j)})\non\\
& &=\sum_{\sig\in S_{m+n}} \de(r_1-\ti v_{\sig(1)} )\ldots  \de(r_n-\ti v_{\sig(n)})\de(k_1-\ti v_{\sig(n+1)})\ldots \de(k_m-\ti v_{\sig(n+m)})\\
& &=\sum_{\sig\in S_{m+n}}\bigg(\prod_{(i, \sig(i))\in C_n\times C_{\ti n}} \de(r_i-\ti r_{\sig(i)} )\bigg)
\bigg(\prod_{(i,\sig(i))\in C_n\times C_{\ti n}' } \de(r_i-\ti k_{\sig(i)-\ti n} )\bigg)\non\\
& &\quad\quad\quad \times\bigg(\prod_{(i,\sig(i))\in C_n'\times C_{\ti n} } \de(k_{i-n}-\ti r_{\sig(i)} )\bigg) 
\bigg(\prod_{(i,\sig(i))\in C_n'\times C_{\ti n}' } 
\de(k_{i-n}-\ti k_{\sig(i)-\ti n} )\bigg),
\eeqa
which concludes the proof. \qed
\bel\label{double-creation-lemma} Let $F_{n,m}, F'_{n,m} \in  L^2((\real^3\times\real^{3n})\times (\real^3\times\real^{3m}))$ be symmetric in the photon variables.  Let us introduce the following operators on $\mcC$
\beqa
B^*_{n,m}(F_{n,m}):=\int d^3q d^3p\int d^{3n}r  d^{3m}k\, F_{n,m}(q;r \ba p ; k) \bb^*(r)^{n}\bb^*(k)^m\nn^*(p-\unk)\nn^*(q-\unr)
\label{double-creation-two}
\eeqa
and set $B_{n,m}(F_{n,m}):=(B^*_{n,m}(F_{n,m}))^*$. There holds
\beqa
& &\lan  B_{\ti n,\ti m}(F'_{\ti n,\ti m})^*\vac, B_{n,m}(F_{n,m})^*\vac\ran
=\sum_{\sig\in S_{m+n}}\int  d^3 q d^3p \int  d^{3n}r d^{3m}k\, F_{n,m}(q;r \ba p; k) \non\\
& &\times\bigg( \ov F_{\ti n,\ti m}'(p-\uhk+\uhr; \hr, \chk \ba q+\uhk-\uhr; \hk, \chr )     
+ \ov F_{\ti n,\ti m}'(q+\uchk-\uchr; \hr, \chk \ba p-\uchk+\uchr; \hk, \chr) \bigg)
\eeqa
for $n+m=\ti n+\ti m$. Otherwise the expression on the l.h.s.  is zero. Here $S_{m+n}$ is the set of permutations of an $m+n$ element set
and the notation $\hk, \chk, \hr, \chr$ is explained in Lemma~\ref{combinatorics}.
\eel
\proof We compute the expectation value
\beqa
 & &\lan  B_{\ti n,\ti m}^*(F'_{\ti n,\ti m})\vac, B_{n,m}^*(F_{n,m})\vac\ran \\
&=&\int d^3\ti q  d^3\ti p d^3 q d^3p \int d^{3\ti n}\ti r d^{3\ti m}\ti k   d^{3n}r d^{3m}k\,  
\ov F'_{\ti n,\ti m}(\ti q;\ti r \ba \ti p; \ti k) F_{n,m}(q;r \ba p; k)\non\\   
& &\times\big( \de(\ti q-p+\ti{\unk}-\unr)+\de(\ti q- q-\ti{\unr}+\unr)\big)\de(\ti p+\ti q-p-q)\non\\
& &\times \lan \vac, \bb(\ti r)^{\ti n} \bb(\ti k)^{\ti m} \bb^*(r)^{n} \bb^*(k)^{m}\vac\ran.       
\label{generalized-creation-operators-one}
\eeqa
The last factor is non-zero only if $\ti n+\ti m=n+m$. 
Then
\beqa
\lan\vac, \bb(\ti r)^{\ti n} \bb(\ti k)^{\ti m} \bb^*(r)^{n} \bb^*(k)^{m}\vac\ran
=\sum_{\sig\in S_{m+n}}\de(\hr-\hat{\ti  r})\de(\chr-\check{\ti k})\de(\chk-\check{\ti r})\de(\hk-\hat{\ti k} ),
\eeqa
where we made use of   Lemma~\ref{combinatorics}.
Thus the r.h.s. of (\ref{generalized-creation-operators-one}) is a sum over $\sig\in S_{m+n}$ of terms of the form:
\beqa
& &\int d^3\ti q  d^3\ti p d^3 q d^3p \int  d^{3n}r d^{3m}k\,  
 \ov F'_{\ti n,\ti m}( \ti q; \hr, \chk \ba \ti p; \hk, \chr)    F_{n,m}(q; r \ba p; k )    \non\\ 
 & &\ph{444444444444444}\times\de(\ti p+\ti q-p-q)\big( \de(\ti q-p+\uhk-\uhr)+\de(\ti q- q-\uchk+\uchr) \big)\non\\
&=&\int  d^3 q d^3p \int  d^{3n}r d^{3m}k\,  
 F_{n,m}(q; r \ba p; k )  \bigg(\ov F_{\ti n,\ti m}'(p-\uhk+\uhr; \hr, \chk \ba q+\uhk-\uhr; \hk, \chr ) \non\\
& &\ph{444444444444444444444444444}+ \ov F_{\ti n,\ti m}'( q+\uchk-\uchr; \hr, \chk \ba p-\uchk+\uchr; \hk, \chr )\bigg) 
\eeqa   
which concludes the proof. \qed

\bel\label{norms-of-scattering-states} Let $G_{1,m},G_{1,m}',G_{2,m}, G_{2,m}'\in L^2(\real^3\times \real^{3m})$ be symmetric in the photon variables. 
We define, as an operator on $\mcC$,
\beqa
B_m^*(G_{i,m}):=\int d^3p\,d^{3m}k\, G_{i,m}(p;k) \neta^*(p-\unk)a^*(k)^m
\eeqa
and set $B_m(G_{i,m})=(B_m^*(G_{i,m}))^*$. There holds the identity
\beqa
\lan\vac, B_{\ti n}(G_{1,\ti n}' )B_{\ti m}(G_{2,\ti m}' )B_{n}^*(G_{1,n}) B_{m}^*(G_{2,m})\vac\ran\pha{444444444444444444444444444444}\non\\
=\sum_{\sig\in S_{m+n}}\int  d^3 q d^3p \int  d^{3n}r d^{3m}k\,  
G_{1,n}(q;  r) G_{2,m}(p;  k)\pha{44444444444444444444444}\non\\
 \times\bigg(\ov G_{1,\ti n}'(p-\uhk+\uhr; \hr, \chk) \ov G_{2,\ti m}'(q+\uhk-\uhr; \hk, \chr)
+\ov G_{1,\ti n}'( q+\uchk-\uchr ;  \hr, \chk)
\ov  G_{2,\ti m}'(p-\uchk+\uchr; \hk, \chr)\bigg),
\eeqa
for $n+m=\ti n+\ti m$. Otherwise the expression on the l.h.s.  is zero. Here $S_{m+n}$ is the set of permutations of an $m+n$ element set
and the notation $\hk, \chk, \hr, \chr$ is explained in Lemma~\ref{combinatorics}.
\eel
\proof Follows immediately from Lemma~\ref{double-creation-lemma}. \qed

\bel\label{norms-of-scattering-states-one} Let $G_{1,m},G_{2,m} \in L^2(\real^3\times \real^{3m})$ be supported in 
$\real^3\times \{\, k\in \real^3\,|\, |k|\geq \si\,\}^{\times m}$ and symmetric in their photon variables.  
There holds the identity
\beqa
& &\lan\vac, B_{\tin}(\Gtin ) (\cH_{\I,\si}^{\cin} )^*  B_{\tim}(\Htim )B_{n}^*(\Gn) \cH_{\I,\si}^{\cin}  B_{m}^*(\Hm)\vac\ran\non\\
&=&\sum_{\sig\in S_{m+n}}\int   d^3 q d^3p \int    d^{3n}r d^{3m}k\, \Gn(q;r) \Hm(p; k)\non\\
& &\times\bigg( \int d^3\ti p\, \cvv^{\si}(\ti p)^2  \ovGtin (p-\ti p-\uhk+\uhr;\hr,\chk) \ovHtim(\ti p+q+\uhk-\uhr; \hk,\chr)\non\\ 
& &\pha{4444444444444444444}+\|\cvv^{\si}\|_2^2  \ovGtin(q+\un{\chk}-\un{\chr};\hr,\chk) 
\ovHtim(p-\un{\chk}+\un{\chr}; \hk,\chr) \bigg)  
\label{H-check-auxiliary}
\eeqa
for $m+n=\tim+\tin$, otherwise the l.h.s. is zero.  Here $S_{m+n}$ is the set of permutations of an $m+n$ element set
and the notation $\hk, \chk, \hr, \chr$ is explained in Lemma~\ref{combinatorics}.
\eel
\proof We compute the expectation value
\beqa
& &\lan\vac, B_{\tin}(\Gtin ) (\cH_{\I,\si}^{\cin} )^*  B_{\tim}(\Htim )B_{n}^*(\Gn)  \cH_{\I,\si}^{\cin}  B_{m}^*(\Hm)\vac\ran\non\\  
&=&\int d^3\ti u d^3\ti w d^3u d^3w\int d^3\ti q  d^3\ti p d^3 q d^3p \int d^{3\tin}\ti r d^{3\tim}\ti k   d^{3n}r d^{3m}k\, \non\\ 
& & \cvv^{\si}(\ti w) \cvv^{\si}(w)\ovGtin(\ti q;\ti r) \ovHtim(\ti p; \ti k)  \Gn(q;r) \Hm(p; k)\non\\ 
& &\times\lan \vac,  \neta(\ti p-\ti{\unk})  \nn^*(\ti u) \nn(\ti u-\ti w)   \neta(\ti q-\ti{\unr}) \neta^*(q-\unr)\nn^*(u-w)\nn(u)\neta^*(p-\unk) \vac\ran\non\\
& &\times\lan \vac, \bb(\ti r)^{\tin}\bb(\ti w) \bb(\ti k)^{\tim} \bb^*(r)^{n} \bb^*(w) \bb^*(k)^{m}\vac\ran.       
\label{generalized-creation-operators}
\eeqa
We note that
\beqa
& &\lan \vac,  \neta(\ti p-\ti{\unk})  \nn^*(\ti u) \nn(\ti u-\ti w)   \neta(\ti q-\ti{\unr})      \neta^*(q-\unr)  \nn^*(u-w)\nn(u)    
 \neta^*(p-\unk) \vac\ran\non\\
&=&\de(\ti p-\ti{\unk}-\ti u)\de(p-\unk- u)\lan \vac,\nn(\ti u-\ti w) \neta(\ti q-\ti{\unr}) \neta^*(q-\unr)\nn^*(u-w) \vac\ran\non\\
&=&\de(\ti p-\ti{\unk}-\ti u)\de(p-\unk- u)\bigg( \de(\ti u-\ti w-q+\unr)\de( \ti q-\ti{\unr}-u+w)\non\\
& &\pha{444444444444444444444444444}+\de(\ti u-\ti w-u+w)\de(\ti q-\ti{\unr}-q+\unr) \bigg).\label{deltas-with-insertions}
\eeqa
Let us consider the contribution to (\ref{generalized-creation-operators}) of the first term in the bracket in (\ref{deltas-with-insertions}):
\beqa
& &\lan\vac, B_{\tin}(\Gtin ) (\cH_{\I,\si}^{\cin}  )^*  B_{\tim}(\Htim )B_{n}^*(\Gn) \cH_{\I,\si}^{\cin}  B_{m}^*(\Hm)\vac\ran_1\non\\
&:=&\int d^3\ti u d^3\ti w d^3u d^3w\int d^3\ti q  d^3\ti p d^3 q d^3p \int d^{3\tin}\ti r d^{3\tim}\ti k   d^{3n}r d^{3m}k\non\\ 
& & \cvv^{\si}(\ti w) \cvv^{\si}(w)\ovGtin(\ti q;\ti r) \ovHtim(\ti p; \ti k)  \Gn(q;r) \Hm(p; k)\non\\
& &\times\de(\ti p-\ti{\unk}-\ti u)\de(p-\unk- u) \de(\ti u-\ti w-q+\unr)\de( \ti q-\ti{\unr}-u+w)\non\\
& &\times\lan \vac, \bb(\ti r)^{\tin}\bb(\ti w) \bb(\ti k)^{\tim} \bb^*(r)^{n} \bb^*(w) \bb^*(k)^{m}\vac\ran\non\\
&=&\int d^3\ti q  d^3\ti p d^3 q d^3p \int d^{3\tin}\ti r d^{3\tim}\ti k   d^{3n}r d^{3m}k\,  \non\\
& &\cvv^{\si}(\ti w_*) \cvv^{\si}(w_*)\ovGtin(\ti q;\ti r) \ovHtim(\ti p; \ti k)  \Gn(q;r) \Hm(p; k)\non\\
& &\times \lan \vac, \bb(\ti r)^{\tin} \bb(\ti k)^{\tim} \bb(\ti w_* )\bb^*(w_*) \bb^*(r)^{n}  \bb^*(k)^{m}\vac\ran, \label{first-term-from-n-deltas}
\eeqa
where in the last step we integrated over $u, w, \ti u, \ti w$ and set
$w_*:=p-\unk-\ti q+\ti{\unr}$, $\ti w_*:=\ti p-\ti{\unk}-q+\unr$. 
Now we consider the expectation value of the photon creation operators:
\beqa
\lan \vac, \bb(\ti r)^{\ti n} \bb(\ti k)^{\ti m} \bb(\ti w_*)\bb^*(w_*) \bb^*(r)^{n}  \bb^*(k)^{m}\vac\ran
=\de(w_*-\ti w_*)\lan \vac, \bb(\ti r)^{\ti n} \bb(\ti k)^{\ti m} \bb^*(r)^{n}  \bb^*(k)^{m}\vac\ran\label{many-deltas},
\eeqa
for $r,k,w, \ti r,\ti k,\ti w$ in the supports of the respective functions. (Here we made use of the fact that $|\ti w_*|\leq \si$, 
whereas $|r_i|\geq \si$, $|k_j|\geq \si$).
Let us now substitute  the r.h.s. of (\ref{many-deltas}) to (\ref{first-term-from-n-deltas}). 
Making use of Lemma~\ref{combinatorics}, we obtain
\beqa
& &\lan\vac, B_{\tin}(\Gtin ) ( \cH_{\I,\si}^{\cin})^*  B_{\tim}(\Htim )B_{n}^*(\Gn)  \cH_{\I,\si}^{\cin}  B_{m}^*(\Hm)\vac\ran_{1}\non\\
&=&\sum_{\sig\in S_{m+n}}\int d^3\ti q  d^3\ti p d^3 q d^3p \int d^{3\tin}\ti r d^{3\tim}\ti k   d^{3n}r d^{3m}k\,  \non\\
& &\cvv^{\si}(\ti p-\ti{\unk}-q+\unr ) \cvv^{\si}(p-\unk-\ti q+\ti{\unr})
 \ovGtin(\ti q ; \ti r) \ovHtim(\ti p; \ti k)  \Gn(q ;r) \Hm(p; k)\non\\
& &\times\de(p+q-\ti p-\ti q) \de(\hat{\ti{r}}-\hat r )\de(\check{\ti{r}}-\check k)\de(\hat{\ti{k}}-\hat k )
\de(\check{\ti{k}}-\check r).
\eeqa
 By integrating over $\ti q,\ti r,\ti k$, we obtain
\beqa
& &\lan\vac, B_{\tin}(\Gtin ) ( \cH_{\I,\si}^{\cin} )^*  B_{\tim}(\Htim )B_{n}^*(\Gn) \cH_{\I,\si}^{\cin}  B_{m}^*(\Hm)\vac\ran_{1}\non\\
&=&\sum_{\sig\in S_{m+n}}\int  d^3\ti p d^3 q d^3p \int    d^{3n}r d^{3m}k\,  
\cvv^{\si}(\ti p- q -\uhk+\uhr)^2 \non\\
& &\pha{444}\times\ovGtin(p+q-\ti p;\hr,\chk) \ovHtim(\ti p; \hk,\chr)  \Gn(q;r) \Hm(p; k)\non\\
&=&\sum_{\sig\in S_{m+n}}\int  d^3\ti p d^3 q d^3p \int    d^{3n}r d^{3m}k\,  
\cvv^{\si}(\ti p)^2  \non\\
& &\pha{444}\times\ovGtin(p-\ti p-\uhk+\uhr;\hr,\chk) \ovHtim(\ti p+q+\uhk-\uhr; \hk,\chr)  \Gn(q;r) \Hm(p; k),
\eeqa
where in the last step we made a change of variables $\ti p \to \ti p+q+\uhk-\uhr$. This
gives the first term on the r.h.s. of (\ref{H-check-auxiliary}). 

Let us now consider the contribution of the second term  in the bracket on the r.h.s. of formula~(\ref{deltas-with-insertions}):
\beqa
& &\lan\vac, B_{\tin}(\Gtin) (\cH_{\I,\si}^{\cin}  )^*  B_{\tim}(\Htim )B_{n}^*(\Gn) \cH_{\I,\si}^{\cin}  B_{m}^*(\Hm)\vac\ran_2\non\\  
&:=&\int d^3w\int d^3\ti q  d^3\ti p d^3 q d^3p \int d^{3\tin}\ti r d^{3\tim}\ti k   d^{3n}r d^{3m}k\,  
\cvv^{\si}(\ti p-\ti{\unk}-p+\unk+w) \cvv^{\si}(w)\non\\
& &\times \ovGtin(q+\ti{\unr}-\unr;\ti r) \ovHtim(\ti p; \ti k)  \Gn(q;r) \Hm(p; k)\non\\
& &\times\lan \vac, \bb(\ti r)^{\tin}\bb(\ti p-\ti{\unk}-p+\unk+w) \bb(\ti k)^{\tim} \bb^*(r)^{n} \bb^*(w) \bb^*(k)^{m}\vac\ran \non\\
&=& \int d^3w\int  d^3\ti p d^3 q d^3p \int d^{3\tin}\ti r d^{3\tim}\ti k   d^{3n}r d^{3m}k\,  
\cvv^{\si}(\ti p-\ti{\unk}-p+\unk+w) \cvv^{\si}(w)\non\\
& &\times \ovGtin(q+\ti{\unr}-\unr;\ti r) \ovHtim(\ti p; \ti k)  \Gn(q;r) \Hm(p; k)\de(\ti p-\ti{\unk}-p+\unk)\non\\
& &\times\lan \vac, \bb(\ti r)^{\tin}\bb(\ti k)^{\tim} \bb^*(r)^{n} \bb^*(k)^{m}\vac\ran \non\\
&=& \|\cvv^{\si}\|_2^2 \sum_{\sig\in S_{m+n}}   \int   d^3 q d^3p \int   d^{3n}r d^{3m}k\,  \Gn(q;r) \Hm(p; k) \non\\
& &\pha{44444444444444444444444444}\times \ovGtin(q+\un{\chk}-\un{\chr};\hr,\chk) \ovHtim(p-\un{\chk}+\un{\chr}; \hk,\chr),
\eeqa
where in the first step we integrated over $\ti u, \ti w, u$ and in the last step  we made use again of Lemma~\ref{combinatorics}. This gives the second term on the r.h.s. of (\ref{H-check-auxiliary}) and concludes the proof.  \qed 

\end{document}